\documentclass[prd,aps,nofootinbib,notitlepage,showpacs,showkeys,preprintnumbers]{revtex4-1}
\usepackage{graphicx,epsf,amsmath,amsfonts,amssymb,amsbsy}
\usepackage{epsfig}
\usepackage{epstopdf}
\usepackage{tikz}
\usepackage{verbatim}

\newcommand{\ds}{\displaystyle}
\newcommand{\vev}[1]{\langle#1\rangle}
\newcommand{\mat}{\left ( \begin{array}}
\newcommand{\emat}{\end{array} \right )}
\newcommand{\vect}{\left ( \begin{array}{c}}
\newcommand{\evect}{\end{array} \right )}

\newcommand{\Det}{\mathop{\rm Det}\nolimits}

\begin{document}

\title{ \bf Dualities in dense quark matter with isospin, chiral and chiral isospin imbalance in the framework of the large-$N_c$ limit of the NJL$_4$ model}
\author{T. G. Khunjua$^{1}$,  K. G. Klimenko$^{2}$, R. N. Zhokhov$^{3}$}
\vspace{1cm}
\affiliation{$^{1)}$ Faculty of Physics, Moscow State University,
119991, Moscow, Russia}
 \affiliation{$^{2)}$ Logunov Institute for High Energy Physics,
NRC "Kurchatov Institute", 142281, Protvino, Moscow Region, Russia}
\affiliation{$^{3)}$  Pushkov Institute of Terrestrial Magnetism, Ionosphere and Radiowave Propagation (IZMIRAN),
108840 Troitsk, Moscow, Russia}

\begin{abstract}
In this paper the phase structure of the dense quark matter has been investigated in the presence of baryon $\mu_B$, isospin $\mu_I$, chiral $\mu_{5}$ and chiral isospin $\mu_{I5}$  chemical potentials in the framework of Nambu--Jona-Lasinio model. It has been shown that in the large-$N_{c}$ limit ($N_{c}$ is the number of quark colours) there exist three duality correspondences in the model. The first duality is between the chiral symmetry breaking and the charged pion condensation phenomena. And there are two new dualities that hold only for chiral symmetry breaking and charged pion condensation phenomena separately.
These dualities show that chiral symmetry breaking phenomenon does not feel the difference between chiral $\mu_{5}$ and chiral isospin $\mu_{I5}$ chemical potentials and charged pion condensation phenomenon does not feel the difference between isospin $\mu_{I}$ and chiral $\mu_{5}$ chemical potentials. It was shown that $\mu_{5}$ can generate charged pion condensation, but this generation occurs at not so large baryon densities. In the case of both chiral imbalances (chiral $\mu_{5}$ and chiral isospin $\mu_{I5}$ chemical potentials) the phase portrait is rather rich, and charged pion condensation in dense quark matter is shown to take up a large part of the phase diagram.
Charged pion condensation in dense quark matter happens even in the case of zero isospin imbalance and requires only chiral imbalances, this fact can be demonstrated with use of one of the new dualities and this is only one example when these dualities are of the great use in exploring the phase diagram.
\end{abstract}


\maketitle

\section{Introduction}
It is widely believed that Quantum Chromodynamics (QCD) is the theory that describes strong interacting processes that occur in dense and/or hot baryonic (quark) matter, which is formed in neutron stars, in heavy ion collision experiments or was created in Early Universe, etc. It is clear that first principle perturbative QCD method can be applied for analytical investigation of these processes only in the high energy region. But at low energies nonperturbative QCD methods (like large-$N_c$ expansion, lattice QCD, etc) and/or different effective theories (models) are usually used for consideration of QCD phenomena.

Each effective model can be regarded only as a more or less plausible approximation of low energy QCD, so this is not a first principle approach. On the other side, numerical simulations of QCD on the lattice (the method is called lattice QCD), turn out to be a real nonperturbative approach to the theory, starting from first principles. But the region of non-vanishing (rather large) baryon chemical potential $\mu_B$ remains out of reach of present lattice computations, due to the famous fermion sign problem (complex fermion determinant), because the main method of lattice QCD -- Monte Carlo simulations -- cannot be applied to QCD at finite $\mu_B$. So dense quark matter is studied, as a rule, in terms of effective field theories. The most widely used low energy effective model for QCD is Nambu--Jona-Lasinio (NJL) model \cite{Nambu:1961fr} (see for review Refs  \cite{Klevansky:1992qe,Hatsuda:1994pi,Buballa:2003qv}). The degrees of freedom in this model are not hadrons but quarks. They are self-interacting and there are no gluons in considerations, they are in a sense integrated out. The most attractive feature of NJL models is the dynamical breaking of the chiral symmetry (quark acquirement of a comparatively large mass) and hence it can be used as a basis model for constituent quark model. But the main drawback of NJL model is a lack of confinement. However one can extend NJL model to the so-called Polykov NJL model (PNJL model, see, e.g., in Ref.  \cite{Fukushima:2003fw}), in which confinement is imitated by a background (temporal) gauge field representing Polyakov loop dynamics.

Here we use the NJL model and try to capture the physics of chiral symmetry breaking and other phenomena, 
which can be rather properly described in terms of this model.
Of course, there is no confinement in our consideration and it is quite a drawback but the qualitative description of different
phenomena can be attained even in the framework of NJL model.
Description in terms of PNJL model can grasp the confinement/deconfinement phase transition and it is a much better approximation for real QCD. It would add the completely new rich structure of confinement/deconfinement regimes but it is rather involved consideration and it should not change qualitatively the results of this paper, probably shifts slightly some phase transition lines. Let us note that the NJL model does not have the sign problem because it does not have gluons at all. But the PNJL model, which capture the gluon contribution and the confinement, has the sign problem just like QCD but it is not that severe \cite{Fukushima:2017csk}.

In addition to temperature and baryon chemical potential $\mu_B$, there are other quantities that describe real quark matter (note that further on we discuss quark matter composed of only light $u$ and $d$ quarks). One of them is the isospin chemical potential $\mu_I$, which is involved in considering systems with isospin asymmetry (different densities $n_u$ and $n_d$ of $u$ and $d$ quarks, correspondingly). Isospin asymmetry does exist in nature, for example, in the case of neutron stars.
Another example is relativistic heavy-ion collisions, where isospin asymmetry is a very plausible scenario. Recently, it was shown that the charged pion condensation (PC) phase is generated in QCD matter if $\mu_{I}$ is greater than the pion mass. This result was obtained in the framework of an effective chiral Lagrangians approach \cite{Son:2000xc,VillavicencioReyes:2004pq,Loewe:2005yn,Splittorff:2000mm} and was supported by QCD lattice calculations, performed at zero
baryon chemical potential $\mu_{B}$ \cite{Brandt:2016zdy, Gupta:2002kp,KogutSinclair}.

From a long time ago the idea of a pion condensate in the core of neutron stars has been considered in connection with the cooling process of a neutron star (see, e.g., Refs \cite{Yakovlev:2004iq}). This idea is one of the motivations to study the behavior of pions at extreme both isospin and baryon densities, searching for phase transitions in the hadronic matter, etc. The generation of charged PC phenomenon was also found in the NJL model \cite{son,eklim,andersen,ak}, but the existence of the charged PC phase in dense quark matter with isospin imbalance was predicted there without sufficient certainty. Then the factors that can promote this generation have been found. It was shown that a charged PC phase might be realized in dense quark system with finite size \cite{ekkz}, in the case of a spatially inhomogeneous pion condensate \cite{gkkz}, in the case of chiral isospin imbalance in the system \cite{ekk,kkz}, or by a rotation in magnetic field \cite{liu}. Some conclusions have been made using a (1+1)-dimensional
toy NJL model, some have been shown to be more general and proved to be to some extent model independent (influence of chiral isospin imbalance \cite{ekk,kkz}).

There is another captivating phenomenon that falls into spotlight quite recently, it is the chiral imbalance (different densities of right-handed $n_R$ and left-handed $n_L$ quarks). \footnote{The distinction between chiral isospin imbalance taken into account in Refs \cite{ekk,kkz} and chiral imbalance phenomenon will be clarified below as well as in Sec. II.} The study of how chiral imbalance can influence the phase diagram of QCD is now getting an increasing attention. This phenomenon stems from the highly nontrivial nature of the vacuum of non-Abelian gauge theories in general,
and of QCD in particular, that allows for the existence of topological
solutions like sphalerons. Sphalerons are classical solutions describing
transitions going above the barrier between the vacua. Sphaleron processes are allowed at high temperatures and, through the Adler-Bell-Jackiw (chiral) anomaly, in the framework of QCD they can generate a chiral imbalance.
It is expected to occur in event-by-event P and CP violating processes in heavy-ion collisions \cite{Kharzeev:2007jp}. In addition, media with chiral imbalance (chiral media) can be realized in Dirac and Weyl semimetals \cite{Li:2014bha}, in Early Universe \cite{Mukherjee:2017bza,Jiang:2014ura}, in neutron stars and supernovae \cite{Charbonneau:2009ax,CharbonneauHoffman}, i.e. in various physical systems and it is important to study their properties.

It is clear from what was said above that chiral imbalance of the system is characterized by the quantity $n_{5}=n_{R}-n_{L}$, which is usually called chiral charge density. In the grand canonical approach to QCD instead of $n_{5}$ the corresponding chiral chemical potential $\mu_5$ appears. In a microscopic picture the chiral charge $n_{5}$ is more relevant quantity, but for technical reasons it is easier to work with $\mu_{5}$. Due to finite quark condensate (which is responsible for processes that change chirality) as well as quantum chiral anomaly, the chiral charge $n_{5}$ is not a strictly conserved quantity. Therefore, $\mu_{5}$ chemical potential is not conjugated to a strictly conserved quantity. Denoting by $\tau$ the typical time scale in which chirality changing processes take place, one can treat $\mu_{5}$ as the chemical potential that describes a system in thermodynamical equilibrium with a fixed value of $n_{5}$ on a time scale much larger than $\tau$ (on the time scale that larger than the one needed for $n_{5}$ to equilibrate).

Earlier, the influence of the chiral $\mu_{5}$ chemical potential on the properties of quark matter was investigated at $\mu_I=0$ in the framework of some effective theories \cite{andrianov,braguta,braguta2,AndrianovEspriu1,Farias:2016let,Ruggieri:2011xc,Ruggieri:2011wd,Frasca:2016rsi}.
Since baryon charge, isospin asymmetry and chiral charge are possible physical parameters of real quark matter, an interesting task how the competition between chemical potentials $\mu_B,\mu_I$ and $\mu_5$ influences its phase structure is in order. In particular, we are interested in solving the question whether (and if it does, under what conditions) the chiral chemical potential $\mu_5$ promotes the generation of a charged PC phase in dense quark matter. Taking into account all these questions, we will however further extend our approach to a problem by the following way.

Note that the notion ``chiral charge density`` $n_5$ can be introduced also for individual quark flavors. Namely, $n_{u5}\equiv n_{uR}-n_{uL}$ and $n_{d5}\equiv n_{dR}-n_{dL}$ are chiral charge densities of $u$ and $d$ quarks, respectively. It is evident that $n_5=n_{u5}+n_{d5}$. Then it is possible to consider the quantity $n_{I5}\equiv (n_{u5}-n_{d5})/2$, which is called chiral isospin charge of the system, as well as the corresponding chiral isospin chemical potential $\mu_{I5}$. In contrast to chiral charge $n_5$, the chiral isospin charge density $n_{I5}$ is a conserved quantity in simplest NJL models, which describe systems composed of light $u$ and $d$ quarks. However, since gluons interact in the same way with different light-quark flavors, it is usually supposed that in two-flavored QCD chiral charges $n_{u5}$ and $n_{d5}$ are equal, hence for real QCD quark matter we usually have $n_{I5}=0$ and $\mu_{I5}=0$. Nevertheless, in the present paper we will study in the large-$N_c$ limit a phase structure of the two-flavored NJL model which is extended by four different chemical potentials, $\mu_{B}$, $\mu_{I}$, $\mu_{5}$, and $\mu_{I5}$. 

Previously, chiral imbalance in the form of chiral isospin ($\mu_{I5}$) chemical potential was considered both in quark and pionic media in the framework of different model approaches \cite{Hanada:2011jb,ekk,kkz,zakharov}. In particular, it was shown in Refs  \cite{ekk,kkz} that in the large-$N_c$ limit ($N_c$ is the number of colors) there is a duality between chiral symmetry breaking (CSB) and charged PC phenomena at $\mu_5=0$. It means that the phase portrait of the model under consideration obeys a symmetry with respect to simultaneous transformations, CSB$\leftrightarrow$charged PC and $\mu_{I}\leftrightarrow\mu_{I5}$. One of the goals of the present paper is to understand how two different kinds of chiral asymmetry, characterized by chemical potentials $\mu_5$ and $\mu_{I5}$, respectively, interplay and influence the phase structure of the NJL model and, especially, the duality between CSB and charged PC phases. In our paper we show that the inclusion of yet another chiral chemical potential ($\mu_5$) greatly enriched the duality properties of the phase portrait of the model. Furthermore, at $\mu_I\ne 0$ the chiral chemical potential $\mu_5$ can induce charged PC phase in dense quark matter.

In the QCD 
there could be more complicated light meson condensations such as rho-meson, $\omega$ meson, kaon etc. that are not considered in our paper and in principle can break the duality of the phase structure. Let us make a couple of comments on them. 
It was suggested early on that at sufficiently high $\mu_{I}$, charged $\rho$-mesons will undergo Bose-Einstein condensation as pions \cite{Lenaghan:2001sd,Sannino:2002wp}. In terms of the holographic model for QCD at nonzero isospin density
in Ref.  \cite{Aharony:2007uu} it has been shown that, indeed, $\rho$ mesons condense for sufficiently high values of $\mu_{I}$ ($\mu_{I} > 1.7m_{\rho}$).
This would have far-reaching consequences for the structure of isospin-rich nuclear matter but it has been concluded
in \cite{Brauner:2016lkh} that $\rho$-meson condensation is either avoided or postponed to isospin chemical potentials much higher than the $\rho$-meson mass. In the context of our consideration even simple argument based on a naive estimate based on the vacuum mass $m_{\rho}$ let alone above-mentioned studies \cite{Brauner:2016lkh} is outside of the parameter range that we are interested in (range of validity of NJL model). So in our consideration one can omit the possibility of $\rho$-meson condensation and the corresponding interaction. 

It is well-known from the Walecka model \cite{Walecka:1974qa}, that vector channel is quite important at nonzero densities. The inclusion of the vector interaction (and temporal vector $\omega$-meson mode, $\omega_0 = q^\dag q$, condensation at finite chemical potential $\mu_B$) in the NJL type models was discussed, e.g., in \cite{Hatsuda:1994pi,Buballa:2003qv,Buballa:1996tm,Sakai:2008ga,Friesen:2014mha}. Vector interaction and $\omega$-meson condensation play an important role in the stabilization of quark matter, as well as in the appearance of a spatially inhomogeneous phases on the QCD phase diagram \cite{Buballa:1996tm}. Note also that the position of the critical endpoint on the QCD phase diagram is still under debate and when the vector interaction is taken into account, the first order transition line decreases in length, the critical endpoint appears at a higher chemical potential $\mu_B$ and lower temperature and can be even removed from the phase diagram \cite{Friesen:2014mha}. However, due to the fact that the strength of this interaction is unknown and, as a rule, it shifts chiral/deconfinement transition line along the chemical potential axis only by maximum of several tens of MeV for rather large values of the vector coupling constant, hence it would not qualitatively change the results of our invesigation. Moreover, since the nonzero  $\omega$-meson condensate shifts effectively the quark-number chemical potential value, it does not spoil the duality relations observed in our paper. 
So we assume, for simplicity, that the vector coupling constant is  zero and $\omega$ condensate can not affect the system. 

The paper is organized as follows.
In Sec. II a (3+1)-dimensional NJL model with two massless quark flavors ($u$ and $d$ quarks) that includes four kinds of chemical potentials, $\mu_B,\mu_I,\mu_{I5},\mu_{5}$, is introduced. Furthermore, the symmetries of the model are discussed and its thermodynamic potential (TDP) is presented in the leading order of the large-$N_c$ expansion.
In Sec. III the duality properties (dual symmetries) of the model TDP are established. Each duality property of the model means that its TDP is invariant under some interchange of chemical potentials as well as, in some cases, simultaneous interchange of condensates. The expressions for the TDP and particle densities in different phases are obtained in the section. Section IV contains the discussion on the phase structure of the model and its different phase portraits are depicted at zero temperature. Moreover, here the role of the duality between CSB  and charged PC phenomena and its influence on the phase diagram is explained (Sec. IV A). Also, other dualities valid only for one of the chiral symmetry breaking or charged pion condensation phenomena are discussed here. In Sec. IV B the case of $\mu_{I5}=0$ and $\mu_5\ne 0$ is considered. Here we show that chiral imbalance ($\mu_5\ne 0$) promotes the appearance of the charged PC phase in dense quark matter with isospin asymmetry. Section IV C contains the consideration of the general case when both chiral isospin $\mu_{I5}$ and chiral $\mu_5$ chemical potentials are nonzero. In Sec. V summary and conclusions are given. Some technical
details are relegated to Appendix A.

\section{The model}

We study a phase structure of the two flavored (3+1)-dimensional NJL model with several chemical potentials. Its Lagrangian, which is symmetrical under global color $SU(N_c)$ group, has the form
\begin{eqnarray}
&&  L=\bar q\Big [\gamma^\nu\mathrm{i}\partial_\nu
+\frac{\mu_B}{3}\gamma^0+\frac{\mu_I}2 \tau_3\gamma^0+\frac{\mu_{I5}}2\tau_{3} \gamma^0\gamma^5+\mu_{5} \gamma^0\gamma^5\Big ]q+ \frac
{G}{N_c}\Big [(\bar qq)^2+(\bar q\mathrm{i}\gamma^5\vec\tau q)^2 \Big
]  \label{1}
\end{eqnarray}
and describes dense baryonic matter with two massless $u$ and $d$ quarks, i.e. $q$ in (1) is the flavor doublet, $q=(q_u,q_d)^T$, where $q_u$ and $q_d$ are four-component Dirac spinors as well as color $N_c$-plets (the summation in (\ref{1}) over flavor, color, and spinor indices is implied); $\tau_k$ ($k=1,2,3$) are Pauli matrices. The Lagrangian (1) contains baryon $\mu_B$-, isospin $\mu_I$-, chiral isospin $\mu_{I5}$-, and chiral $\mu_{5}$ chemical potentials. In other words, this model is able to describe the properties of quark matter with nonzero baryon $n_B=(n_{u}+n_{d})/3\equiv n/3$, isospin $n_I=(n_{u}-n_{d})/2$, chiral isospin $n_{I5}=(n_{u5}-n_{d5})/2$ and chiral $n_{5}=n_{R}-n_{L}$ densities which are the quantities, thermodynamically conjugated to chemical potentials $\mu_B$, $\mu_I$, $\mu_{I5}$ and $\mu_{5}$, respectively.  (We use above the notations $n_f$ and $n_{fL(R)}$ for density of quarks as well as density of left(right)-handed quarks with indifidual flavor $f=u,d$, respectively. Moreover, $n_{f5}=n_{fR}-n_{fL}$ and $n_{R(L)}=n_{uR(L)}+n_{dR(L)}$.)

The quantities $n_B$, $n_I$ and $n_{I5}$ are densities of conserved charges, which correspond to the invariance of Lagrangian (1) with respect to the abelian $U_B(1)$, $U_{I_3}(1)$ and $U_{AI_3}(1)$
groups, where \footnote{\label{f1,1}
Recall for the following that~~
$\exp (\mathrm{i}\alpha\tau_3)=\cos\alpha
+\mathrm{i}\tau_3\sin\alpha$,~~~~
$\exp (\mathrm{i}\alpha\gamma^5\tau_3)=\cos\alpha
+\mathrm{i}\gamma^5\tau_3\sin\alpha$.}
\begin{eqnarray}
U_B(1):~q\to\exp (\mathrm{i}\alpha/3) q;~
U_{I_3}(1):~q\to\exp (\mathrm{i}\alpha\tau_3/2) q;~
U_{AI_3}(1):~q\to\exp (\mathrm{i}
\alpha\gamma^5\tau_3/2) q.
\label{2001}
\end{eqnarray}
So we have from (\ref{2001}) that $n_B=\vev{\bar q\gamma^0q}/3$, $n_I=\vev{\bar q\gamma^0\tau^3 q}/2$ and $n_{I5}=\vev{\bar q\gamma^0\gamma^5\tau^3 q}/2$. We would like also to remark that, in addition to (\ref{2001}), Lagrangian (1) is invariant with respect to the electromagnetic $U_Q(1)$ group,
\begin{eqnarray}
U_Q(1):~q\to\exp (\mathrm{i}Q\alpha) q,
\label{2002}
\end{eqnarray}
where $Q={\rm diag}(2/3,-1/3)$. However, as it was noted in Introduction, the chiral chemical potential $\mu_5$ does not correspond to a conserved quantity of the model (1). It is usually inroduced in order to describe a system on the time scales, when all chirality changing processes are finished in the system, so it is in the state of thermodynamical equilibrium with some fixed value of the chiral density $n_5$ \cite{andrianov}.

Sometimes the interaction terms in Eq. (1) are written in a more general form,
\begin{eqnarray}
&&  L=\bar q\Big [\gamma^\nu\mathrm{i}\partial_\nu
+\frac{\mu_B}{3}\gamma^0+\frac{\mu_I}2 \tau_3\gamma^0+\frac{\mu_{I5}}2\tau_{3} \gamma^0\gamma^5+\mu_{5} \gamma^0\gamma^5\Big ]q \nonumber\\&&+ \frac
{G_{1}}{N_c}\Big [(\bar qq)^2+(\bar q\mathrm{i}\gamma^5q)^2+(\bar q\vec\tau q)^2 +(\bar q\mathrm{i}\gamma^5\vec\tau q)^2  \Big
]
 + \frac{G_{2}}{N_c}\Big [(\bar qq)^2-(\bar q\mathrm{i}\gamma^5q)^2-(\bar q\vec\tau q)^2 +(\bar q\mathrm{i}\gamma^5\vec\tau q)^2  \Big
]. \label{1G1}
\end{eqnarray}
If $G_{2}=0$ then Lagrangian (\ref{1G1}) has the additional $U_A(1)$ axial symmetry. In this case $n_{5}$ is a conserved charge of the system, which correspond to the invariance of Lagrangian (\ref{1G1}) with respect to the abelian $U_A(1)$ group. In general both $G_{1}\ne 0$ and $G_{2}\ne 0$. The last term in (\ref{1G1}) is 't Hooft's instanton-induced interaction term which breaks explicitly the $U_A(1)$ axial symmetry of  the Lagrangian. In the following, we choose $G_1=G_2\equiv G/2$ in Eq. (\ref{1G1}) and hence study only the standard NJL Lagrangian (1).

The ground state expectation values of $n_B$,  $n_I$, $n_{I5}$ and $n_{5}$ can be found by differentiating the thermodynamic potential of the system (1) with respect to the corresponding chemical potentials. The goal of the present paper is the investigation of the ground state properties (or phase structure) of the system (1) and its dependence on the chemical potentials  $\mu_B$, $\mu_I$, $\mu_{I5}$ and $\mu_{5}$.

To find the TDP of the system, we use a semibosonized version of the Lagrangian (\ref{1}), which contains composite bosonic fields $\sigma (x)$ and $\pi_a (x)$ $(a=1,2,3)$ (in what follows, we use the notations
$\mu\equiv\mu_B/3$, $\nu\equiv\mu_I/2$, $\nu_{5}\equiv\mu_{I5}/2$):
\begin{eqnarray}
\widetilde L\ds &=&\bar q\Big [\gamma^\rho\mathrm{i}\partial_\rho
+\mu\gamma^0+\nu\tau_3\gamma^0+\nu_{5}\tau_{3}\gamma^0\gamma^5+\mu_{5}\gamma^0\gamma^5-\sigma -\mathrm{i}\gamma^5\pi_a\tau_a\Big ]q
 -\frac{N_c}{4G}\Big [\sigma\sigma+\pi_a\pi_a\Big ].
\label{2}
\end{eqnarray}
In (\ref{2}) and below the summation over repeated indices is implied. From the auxiliary Lagrangian (\ref{2}) one gets the equations
for the bosonic fields
\begin{eqnarray}
\sigma(x)=-2\frac G{N_c}(\bar qq);~~~\pi_a (x)=-2\frac G{N_c}(\bar q
\mathrm{i}\gamma^5\tau_a q).
\label{200}
\end{eqnarray}
Note that the composite bosonic field $\pi_3 (x)$ can be identified with the physical $\pi^0(x)$-meson field, whereas the physical $\pi^\pm (x)$-meson fields are the following combinations of the composite fields, $\pi^\pm (x)=(\pi_1 (x)\mp i\pi_2 (x))/\sqrt{2}$.
Obviously, the semibosonized Lagrangian $\widetilde L$ is equivalent to the initial Lagrangian (\ref{1}) when using the equations (\ref{200}).
Furthermore, it is clear from (\ref{2001}) and footnote \ref{f1,1} that the composite bosonic fields (\ref{200}) are transformed under the isospin $U_{I_3}(1)$ and axial isospin $U_{AI_3}(1)$ groups in the following manner:
\begin{eqnarray}
U_{I_3}(1):~&&\sigma\to\sigma;~~\pi_3\to\pi_3;~~\pi_1\to\cos
(\alpha)\pi_1+\sin (\alpha)\pi_2;~~\pi_2\to\cos (\alpha)\pi_2-\sin
(\alpha)\pi_1,\nonumber\\
U_{AI_3}(1):~&&\pi_1\to\pi_1;~~\pi_2\to\pi_2;~~\sigma\to\cos
(\alpha)\sigma+\sin (\alpha)\pi_3;~~\pi_3\to\cos
(\alpha)\pi_3-\sin (\alpha)\sigma.
\label{201}
\end{eqnarray}
Starting from the auxiliary Lagrangian (\ref{2}), one obtains in the leading order of the large-$N_c$ expansion (i.e. in the one-fermion loop
approximation) the following path integral expression for the
effective action ${\cal S}_{\rm {eff}}(\sigma,\pi_a)$ of the bosonic
$\sigma (x)$ and $\pi_a (x)$ fields:
$$
\exp(\mathrm{i}{\cal S}_{\rm {eff}}(\sigma,\pi_a))=
  N'\int[d\bar q][dq]\exp\Bigl(\mathrm{i}\int\widetilde L\,d^4 x\Bigr),
$$
where
\begin{equation}
{\cal S}_{\rm {eff}}
(\sigma(x),\pi_a(x))
=-N_c\int d^4x\left [\frac{\sigma^2+\pi^2_a}{4G}
\right ]+\tilde {\cal S}_{\rm {eff}},\label{3}
\end{equation}
The quark contribution to the effective action, i.e. the term
$\tilde {\cal S}_{\rm {eff}}$ in (\ref{3}), is given by:
\begin{eqnarray}
\exp(\mathrm{i}\tilde {\cal S}_{\rm {eff}})&=&N'\int [d\bar
q][dq]\exp\Bigl(\mathrm{i}\int\Big\{\bar q\big
[\gamma^\rho\mathrm{i}\partial_\rho +\mu\gamma^0+
\nu\tau_3\gamma^0+\nu_{5}\tau_{3}\gamma^0\gamma^5+\mu_5\gamma^0\gamma^5-\sigma -\mathrm{i}\gamma^5\pi_a\tau_a\big
]q\Big\}d^4 x\Bigr)\nonumber\\
&=&[\Det D]^{N_c},
 \label{4}
\end{eqnarray}
where $N'$ is a normalization constant. Moreover, in (\ref{4}) we have introduced the notation $D$,
\begin{equation}
D\equiv\gamma^\nu\mathrm{i}\partial_\nu +\mu\gamma^0
+ \nu\tau_3\gamma^0+\nu_{5}\tau_{3}\gamma^0\gamma^5+\mu_{5}\gamma^0\gamma^5-\sigma (x) -\mathrm{i}\gamma^5\pi_a(x)\tau_a,
\label{5}
\end{equation}
for the Dirac operator, which acts in the flavor-, spinor- as well as coordinate spaces only. Using the general formula $\Det D=\exp {\rm Tr}\ln D$, one obtains for the effective action (\ref{3}) the following expression
\begin{equation}
{\cal S}_{\rm {eff}}(\sigma(x),\pi_a(x))
=-N_c\int
d^4x\left[\frac{\sigma^2(x)+\pi^2_a(x)}{4G}\right]-\mathrm{i}N_c{\rm
Tr}_{sfx}\ln D,
\label{6}
\end{equation}
where the Tr-operation stands for the trace in spinor- ($s$), flavor-
($f$) as well as four-dimensional coordinate- ($x$) spaces, respectively.

The ground state expectation values  $\vev{\sigma(x)}$ and
$\vev{\pi_a(x)}$ of the composite bosonic fields are determined by
the saddle point equations,
\begin{eqnarray}
\frac{\delta {\cal S}_{\rm {eff}}}{\delta\sigma (x)}=0,~~~~~
\frac{\delta {\cal S}_{\rm {eff}}}{\delta\pi_a (x)}=0,~~~~~
\label{05}
\end{eqnarray}
where $a=1,2,3$. Just the knowledge of $\vev{\sigma(x)}$ and
$\vev{\pi_a(x)}$ and, especially, of their behaviour vs chemical potentials supplies us with a phase structure of the model. It is clear from (\ref{201}) that if $\vev{\sigma(x)}\ne 0$ and/or $\vev{\pi_3(x)}\ne 0$, then the axial isospin $U_{AI_3}(1)$ symmetry of the model is spontaneously broken down, whereas at $\vev{\pi_1(x)}\ne 0$ and/or $\vev{\pi_2(x)}\ne 0$ we have a spontaneous breaking of the isospin $U_{I_3}(1)$ symmetry. Since in the last case the ground state expectation values, or condensates, both of the field $\pi^+(x)$ and of the field $\pi^-(x)$ are not zero, this phase is usually called the charged pion condensation (PC) phase. In addition, it is easy to see from (\ref{200}) that the nonzero condensates $\vev{\pi_{1,2}(x)}$ (or $\vev{\pi^\pm(x)}$) are not invariant with respect to the electromagnetic $U_Q(1)$ transformations (\ref{2002}) of the flavor quark doublet. Hence in the charged PC phase the electromagnetic $U_Q(1)$ invariance of the model (1) is also broken  spontaneously, and superconductivity is an unavoidable property of this phase.

In the present paper we suppose that in the ground state of the system, i.e. in the state of thermodynamic equilibrium, the ground state expectation values $\vev{\sigma(x)}$ and $\vev{\pi_a(x)}$ do not depend on spacetime coordinates $x$,
\begin{eqnarray}
\vev{\sigma(x)}\equiv M,~~~\vev{\pi_a(x)}\equiv \pi_a, \label{8}
\end{eqnarray}
where $M$ and $\pi_a$ ($a=1,2,3$) are already constant quantities. In fact, they are coordinates of the global minimum point of the
thermodynamic potential (TDP) $\Omega (M,\pi_a)$.
In the leading order of the large-$N_c$ expansion and using (\ref{8})
it is defined by the following expression:
\begin{equation}
\int d^4x \Omega (M,\pi_a)=-\frac{1}{N_c}{\cal S}_{\rm
{eff}}\big (\sigma(x),\pi_a (x)\big )\Big|_{\sigma
(x)=M,\pi_a(x)=\pi_a} .\label{08}
\end{equation}
In what follows we are going to investigate the
$\mu,\nu,\nu_{5},\mu_{5}$-dependence of the global minimum point of the function $\Omega (M,\pi_a)$ vs $M,\pi_a$. To simplify the task, let us note that due to a $U_{I_3}(1)\times U_{AI_3}(1)$ invariance of the model, the TDP (\ref{08}) depends effectively only on the two combinations, $\sigma^2+\pi_3^2$ and $\pi_1^2+\pi_2^2$, of the bosonic fields, as is easily seen from (\ref{201}). In this case, without loss of generality, one can put $\pi_2=\pi_3=0$ in (\ref{08}),
and study the TDP as a function of only two variables,
$M\equiv\sigma$ and $\Delta\equiv\pi_1$. So, throughout the paper we use the ansatz
\begin{eqnarray}
\vev{\sigma(x)}=M,~~~\vev{\pi_1(x)}=\Delta,~~~\vev{\pi_2(x)}=0,~~~ \vev{\pi_3(x)}=0. \label{06}
\end{eqnarray}
In this case the TDP (\ref{08}) reads
\begin{eqnarray}
\Omega (M,\Delta)~
&&=\frac{M^2+\Delta^2}{4G}+\mathrm{i}\frac{{\rm
Tr}_{sfx}\ln D}{\int d^4x}\nonumber\\
&&=\frac{M^2+\Delta^2}{4G}+\mathrm{i}\int\frac{d^4p}{(2\pi)^4}\ln\Det\overline{D}(p),
\label{07}
\end{eqnarray}
where
\begin{equation}
\overline{D}(p)=\not\!p +\mu\gamma^0
+ \nu\tau_3\gamma^0+\nu_{5}\tau_{3}\gamma^0\gamma^5+ \mu_{5}\gamma^0\gamma^5-M
-\mathrm{i}\gamma^5\Delta\tau_1\equiv\left
(\begin{array}{cc}
A~, & U\\
V~, & B
\end{array}\right )
\label{500}
\end{equation}
is the momentum space representation of the Dirac operator $D$ (\ref{5}) under the constraint (\ref{06}). The quantities $A,B,U,V$ in  (\ref{500}) are really the following 4$\times$4 matrices,
\begin{equation}
A=\not\!p +\mu\gamma^0
+ \nu\gamma^0+\nu_{5}\gamma^0\gamma^5+ \mu_{5}\gamma^0\gamma^5-M;~~B=\not\!p +\mu\gamma^0
- \nu\gamma^0-\nu_{5}\gamma^0\gamma^5 + \mu_{5}\gamma^0\gamma^5-M;~~U=V=-\mathrm{i}\gamma^5\Delta,
\label{80}
\end{equation}
so the quantity $\overline{D}(p)$ from (\ref{500}) is indeed a 8$\times$8 matrix whose determinant appears in the expression (\ref{07}). Based on the following general relations
\begin{eqnarray}
\Det\overline{D}(p)\equiv\det\left
(\begin{array}{cc}
A~, & U\\
V~, & B
\end{array}\right )=\det [-VU+VAV^{-1}B]=\det
[BA-BUB^{-1}V]
\label{9}
\end{eqnarray}
and using any program of analytical calculations, one can find from (\ref{80}) and (\ref{9})
\begin{eqnarray}
\Det\overline{D}(p)=\big (\eta^4-2a_+\eta^2+b_+\eta+c_+\big )\big (\eta^4-2a_-\eta^2+b_-\eta+c_-\big )\equiv P_+(\eta)P_-(\eta),
\label{91}
\end{eqnarray}
where $\eta=p_0+\mu$, $|\vec p|=\sqrt{p_1^2+p_2^2+p_3^2}$ and
\begin{eqnarray}
a_\pm&&=M^2+\Delta^2+(|\vec p|\pm\mu_{5})^2+\nu^2+\nu_{5}^2;~~b_\pm=\pm 8(|\vec p|\pm\mu_{5})\nu\nu_{5};\nonumber\\
c_\pm&&=a_\pm^2-4 \nu ^2
\left(M^2+(|\vec p|\pm\mu_{5})^2\right)-4 \nu_{5}^2 \left(\Delta ^2+(|\vec p|\pm\mu_{5})^2\right)-4\nu^{2} \nu_{5}^2.
\label{10}
\end{eqnarray}
It is clear directly from the relations (\ref{91}) and (\ref{10}) that the TDP (\ref{07}) is an even function over each of the variables $M$ and $\Delta$. Moreover, in the most general case it is invariant under the transformation $\mu\to-\mu$. \footnote{Indeed, if simultaneously with $\mu\to-\mu$ we perform in the integral (\ref{07}) the $p_0\to-p_0$ change of variables, then one can easily see that the expression (\ref{07}) remains intact. } Hence, without loss of generality we can limit ourselves in the following only by $\mu\ge 0$, $M\ge 0$, and $\Delta\ge 0$ values of these quantities. Since in our consideration we study the interplay and competition between two possible types of chiral asymmetry, it is interesting to compare first of all the phase structure of the model (1) in two particular cases, (i) $\nu_5=0$ and (ii) $\mu_5=0$. Then in the case (i), in addition to the previous parity property, the TDP (\ref{07}) is an even function separately with respect to $\nu$, and separately with respect to $\mu_5$, whereas in the case (ii) it is an even function separately over $\nu$, and separately over $\nu_5$. In addition, if $\nu=0$, then the TDP is an even function separately with respect to $\nu_5$, and separately with respect to $\mu_5$. As a result, in all above mentioned cases it is enough to consider only positive values of nonzero chemical potentials, in order to imagine a full phase portrait of the model.

However, if $\nu\ne 0$, $\nu_5\ne 0$ and  $\mu_5\ne 0$, then it is easily seen from relations (\ref{91}) and (\ref{10}) that the TDP (\ref{07}) is symmetric with respect to the following three transformations, in each of them two chemical potentials change their sign simultaneously: $\{\nu\to-\nu;~\nu_5\to-\nu_5\}$, $\{\nu\to-\nu;~\mu_5\to-\mu_5\}$ and $\{\nu_5\to-\nu_5;~\mu_5\to-\mu_5\}$. All these symmetries of the TDP can help to analyze the phase portrait of the model. In particular, at $\nu\ne 0$, $\nu_5\ne 0$ and  $\mu_5\ne 0$ it is sufficient to study the phase structure only in the case, when arbitrary two of $\nu,\nu_5,\mu_5$-chemical potentials take positive values, whereas the sign of the last one is not fixed.

\section{Calculation of the TDP (\ref{07}). Duality relations.}
\subsection{Duality properties of the model}\label{IIIA}

By the duality property (or symmetry, or relation, etc) of any theory, we will understand any symmetry of its TDP with respect to transformations as order parameters (in our case, condensates $M$ and $\Delta$) and free external parameters of the system (these may be chemical potentials, coupling constants, etc). The presence of the dual symmetry of the model means that its phase portrait also has some symmetry with respect to the transformation of external parameters, which can greatly simplify the construction of the phase diagram of the system. (The invariance of the TDP (\ref{07}) under the changing of a sign of its parameters considered at the end of the previous section is the simplest example of the dual symmetry of the model (1). Due to this kind of duality, it is enough to study the phase structure only, e.g., at $\mu\ge 0$, etc.) In general, there might be several duality relations in the system. Below, we consider the most interesting dualities, which exist in the large-$N_c$ limit of the massless NJL model (1).

The first duality property inherent to our model is easily seen from Eqs (\ref{07}), (\ref{91}) and (\ref{10}). Indeed, it is clear from these relations that the TDP of the system is invariant with respect to the transformation
\begin{eqnarray}
{\cal D}:~~~~M\longleftrightarrow \Delta,~~\nu\longleftrightarrow\nu_{5}
 \label{16}
\end{eqnarray}
at fixed values of $\mu$ and $\mu_5$. It is the so-called main duality of the system which means that if at $\mu,\mu_5,\nu=A,\nu_5=B$ the global minimum of the TDP lies at the point $(M=M_0,\Delta=\Delta_0)$, then at $\mu,\mu_5,\nu=B,\nu_5=A$ it is at the point $(M=\Delta_0,\Delta=M_0)$. In the next section we will discuss the influence and meaning of this duality symmetry on the phase structure of the model (1) in more details, but just now we would like to note that at $\mu_5=0$ the property (\ref{16}) leads to the duality between CSB and charged PC phenomena \cite{ekk,kkz}. \footnote{Note that another kind of duality correspondence, the duality between CSB and
superconductivity, was demonstrated both in (1+1)- and
(2+1)-dimensional NJL models \cite{thies,ekkz2}.} In addition to (\ref{16}), in the framework of our model there are two other so-called constrained duality relations, which include into consideration the chemical potential $\mu_5$. To find them let us transform the expression (\ref{91}) for $\Det\overline{D}(p)$. Namely, expanding the polynomials $P_\pm(\eta)$ in a series over $\Delta$, we have
\begin{eqnarray}
P_\pm(\eta)&\equiv&\Delta^4-2\Delta^2\big [\eta^2-(|\vec p|\pm\mu_5)^2-M^2+\nu_5^2-\nu^2\big ]\nonumber\\
&+&\big [M^2+(|\vec p|\pm\mu_5+\nu_5)^2-(\eta\pm\nu)^2\big ]\big [M^2+(|\vec p|\pm\mu_5-\nu_5)^2-
(\eta\mp\nu)^2\big ]. \label{17}
\end{eqnarray}
Then
\begin{eqnarray}
&&\Det\overline{D}(p)\Big |_{\Delta=0}\equiv P_+(\eta)P_-(\eta)\Big |_{\Delta=0}=\big [M^2+(|\vec p|+\mu_5+\nu_5)^2-(\eta+\nu)^2\big ]\nonumber\\
&&
\big [M^2+(|\vec p|+\mu_5-\nu_5)^2-
(\eta-\nu)^2\big ]
\big [M^2+(|\vec p|-\mu_5+\nu_5)^2-(\eta-\nu)^2\big ]\big [M^2+(|\vec p|-\mu_5-\nu_5)^2-(\eta+\nu)^2\big ].
\label{18}
\end{eqnarray}
It follows from Eq. (\ref{18}) that at the constraint $\Delta=0$ the TDP (\ref{07}) is invariant with respect to the transformation
\begin{eqnarray}
{\cal D}_M:~~~~\Delta=0,~~~\mu_5\longleftrightarrow\nu_{5}.
 \label{19}
\end{eqnarray}
In a similar way it is possible to show that
\begin{eqnarray}
&&\Det\overline{D}(p)\Big |_{M=0}\equiv P_+(\eta)P_-(\eta)\Big |_{M=0}=\big [\Delta^2+(|\vec p|+\mu_5+\nu)^2-(\eta+\nu_5)^2\big ]\nonumber\\
&&
\big [\Delta^2+(|\vec p|+\mu_5-\nu)^2-
(\eta-\nu_5)^2\big ]
\big [\Delta^2+(|\vec p|-\mu_5+\nu)^2-(\eta-\nu_5)^2\big ]\big [\Delta^2+(|\vec p|-\mu_5-\nu)^2-(\eta+\nu_5)^2\big ].
\label{20}
\end{eqnarray}
Hence, at the constraint $M=0$ the TDP (\ref{07}) is invariant with respect to the transformation
\begin{eqnarray}
{\cal D}_\Delta:~~~~M=0,~~~\mu_5\longleftrightarrow\nu.
 \label{21}
\end{eqnarray}
Furthermore, one can note that the dualities ${\cal D}_M$ and ${\cal D}_\Delta$ are dual to each other with respect to ${\cal D}$ duality, indeed if one take, for example, ${\cal D}_M$ duality and make the transformation $M\longleftrightarrow \Delta,~~\nu\longleftrightarrow\nu_{5}$, then one gets ${\cal D}_\Delta$ duality and vise versa. So in a sense there exist only one independent additional duality and the other can be obtained by the main duality.
The diagram explaining the dualities and their relations to each other is depicted in Fig. 1.
\begin{figure}
\includegraphics[width=0.7\textwidth]{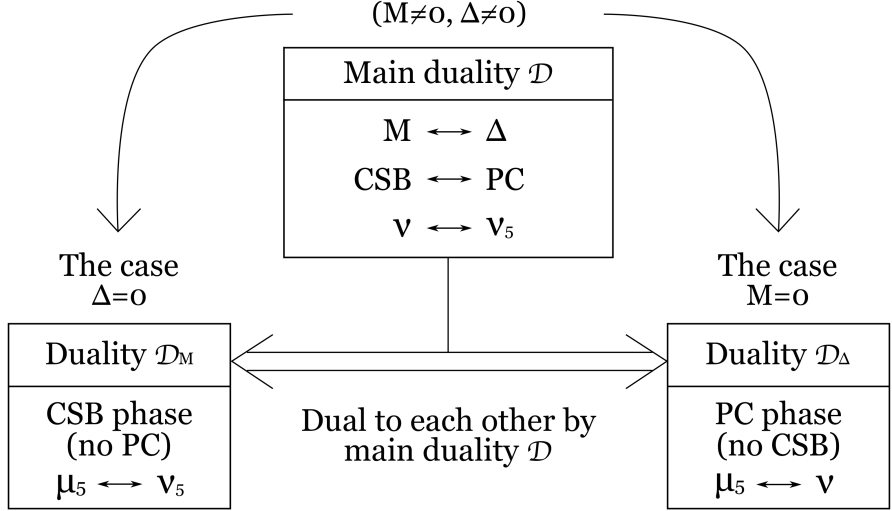}
\caption{ Diagram explaining the duality properties and their
relations. The main duality  $\cal D$ that is valid in the most general case $(M\neq0,\Delta\neq0)$, the dualities  ${\cal D}_{M}$ and ${\cal D}_\Delta$ valid only when $\Delta=0$ and $M=0$,\, i.e. outside of PC and CSB phases.}
\end{figure}

The duality relation (\ref{19}) (the relation (\ref{21})) means that if at some values of the chemical potentials the CSB phase (the charged PC phase) is realized, then at $\mu_5\leftrightarrow\nu_{5}$ (at $\mu_5\leftrightarrow\nu$) the same phase will be observed, if dynamically or due to other reasons the pion condensate $\Delta$ is equal to zero (the chiral  condensate $M$ is equal to zero) in the system.
Let us recall that throughout this paper we consider the phase structure in the chiral limit (zero current quark mass). If one is to consider nonzero current quark mass then 
the main duality ${\cal D}$ and ${\cal D}_\Delta$ duality will be only approximate. But the duality ${\cal D}_M$ will remain exact even without chiral limit.  As it will be shown in the next section, the presence of dual symmetries ${\cal D}$, ${\cal D}_M$ and ${\cal D}_\Delta$ of the model TDP can greatly facilitate the process of finding its phase structure.

Let us elaborate a little more 
on the duality notion. In our paper it is a symmetry relation between condensates (phases) and matter content (chemical potentials). But the notion of duality is more widespread. And it is a very powerful 
concept that is used in different domains of theoretical physics ranging from 
string theory to condensed matter physics etc. For example, there is a class of dualities called strong-weak dualities that connect weak coupling regime of one theory with strong coupling regime of the other. To this class belongs such a famous duality as AdS/CFT (or gauge/gravity) duality \cite{Maldacena:1997re}, which connects some strongly-coupled four-dimensional gauge theories at large $N_{c}$ to tractable weakly-coupled string theories living in ten dimensions.  Now AdS/CFT conjecture is a subject of very intense study. Another example,  which can also be attributed to the strong-weak duality class is the duality between CSB and superconductivity phenomena in low dimensional field theories. Indeed, in this case the weak coupling CSB phenomenon is dually conjugated to strong coupling superconductivity, and vice versa \cite{ekkz2}.

There is another class of dualities, which is also historically connected to AdS/CFT duality. They are called strong-strong dualities or usually bear another name large-$N_{c}$ orbifold equivalences \cite{Hanada:2011jb,Hanada:2011ju,Kashiwa:2017yvy}.
Orbifold equivalences connect gauge theories with different gauge groups and matter content in the large-$N_{c}$ limit. In the framework of orbifold equivalence formalism in \cite{Hanada:2011jb} there have been also obtained a duality between charged PC and chiral symmetry breaking phenomena.
These dualities have been shown only for a large number of colors $N_{c}$, but it was argued that the universality may work approximately even for $N_{c}=3$.

The orbifold equivalence can dually-relate gauge theories with different gauge groups. For example, in the ordinary orbifold equivalence, the sign-problem free gauge theories such as $SO(2N_c)$ and $Sp(2N_c)$ has been used to investigate the QCD phase diagram (outside the charged pion condensation region) that has the sign problem at finite density (at $\mu>0$). This is the big advantage of the dualities via the orbifold equivalence.

Our dualities do not have this advantage because they connect different matter content of the same gauge theory. For example, the QCD phase diagrams ($\mu$, $\mu_{5}$) and ($\mu$, $\mu_{I}$) are connected, but at zero $\mu$ both do not have sign problem. Whereas at non-zero $\mu$ both do. Sign problem is inherent to the QCD with non-zero $\mu$.  The lattice QCD at $\mu_{I5}$ has not been considered so far, and let us not discuss the presence of sign problem in this case. Let us just speculate that if there is a sign problem in the case of $\mu_{I5}$ then it can connect sign problem free QCD with $\mu_{I}$ to the QCD with chiral imbalance, non-zero $\mu_{I5}$ and the sign problem can be circumvented outflanked in this way. If there is a sign problem in the case of $\mu_{I5}$ then one can still use the duality in the following way. If the phase diagram with $\mu$ and one of $\mu_{5}$, $\mu_{I}$, $\mu_{I5}$ (has the sign problem due to non-zero $\mu$) is considered on lattice using any method of lattice QCD at $\mu$ (reweighting, analytic continuation, imaginary chemical potential), for example, ($\mu$, $\mu_{5}$)-phase diagram then the duality can be used to map these results to the other section of the ($\mu$, $\mu_{5}$, $\mu_{I}$, $\mu_{I5}$)-phase diagram, and get, for example, ($\mu$, $\mu_{I5}$)-phase diagram.

Sometimes it is possible getting immediately some nontrivial phase diagrams merely by the duality mapping, for example, we can use the results of NJL model and lattice QCD simulations with non-zero $\mu_{5}$ \cite{braguta, braguta2} and get the QCD phase diagram with non-zero $\mu_{I5}$ and hence establish the catalysis of chiral symmetry breaking by $\mu_{I5}$.

These dualities give us possibly very interesting insights into QCD phase diagram, for instance, the fact that pion condensation phenomenon (in chiral symmetry restored phase) is affected by isospin ($\mu_{I}$) and chiral ($\mu_{5}$) asymmetry in exactly the same way. This might be just a coincidence or maybe there are deep reasons behind that, anyway, it is an interesting feature.

It is well known that the sign problem is absent in four-fermionic theories like the NJL$_4$ model (1), which effectively describes the low energy region of QCD. So, the NJL$_4$ model (1) itself can be used, in principle, for the investigation of the QCD phase diagram at arbitrary permissible values of chemical potentials (less than $\approx 1$ GeV) without using the duality relations ${\cal D}$, ${\cal D}_M$ and ${\cal D}_\Delta$. In our opinion, another not so striking, nevertheless, very pleasant attractive feature of these dualities is the possibility, using the duality mapping, to predict the phase portrait in the dually conjugated region of chemical potentials without spending time on numerical calculations.
Last but not least, we also get the opportunity to use the dual relations to find the values of many physical characteristics of the dually conjugate phases as the values of condensates, baryon density, etc.

Prior to that moment the dualities of the QCD phase diagram have been discussed only in the orbifold equivalence approach and it is nice to establish dualities in other approaches, for example, in this paper the dualities are studied in the framework of effective model (NJL model). These dualities probably can be used to complement the dualities in orbifold equivalence or one can get hints of dualities worth checking in orbifold equivalence.

\subsection{TDP and its projections onto $M$ and $\Delta$ axes. Quark number density}

In order to find the TDP (\ref{19}), we use the following representation for $\Det\overline{D}(p)$ of Eq. (\ref{91})
\begin{eqnarray}
\Det\overline{D}(p)=
\big (\eta-\eta_1\big)\big (\eta-\eta_2\big)\big (\eta-\eta_3\big)\big (\eta-\eta_4\big)\big (\eta-\eta_5\big)\big (\eta-\eta_6\big)\big (\eta-\eta_7\big)\big (\eta-\eta_8\big),
\end{eqnarray}
where half of the eight quantities $\eta_i$ are the roots of the polynomial $P_{+}(\eta)$, and the other half are the roots of the polynomial $P_{-}(\eta)$. So
\begin{eqnarray}
\Omega (M,\Delta)~
=\frac{M^2+\Delta^2}{4G}+\mathrm{i}\sum_{i=1}^{8}\int\frac{d^4p}{(2\pi)^4}\ln(p_{0}+\mu-\eta_{i}).\label{070}
\end{eqnarray}
Then, taking into account a general formula
\begin{eqnarray}
\int_{-\infty}^\infty dp_0\ln\big
(p_0-K)=\mathrm{i}\pi|K|,\label{int}
\end{eqnarray}
one gets
\begin{eqnarray}
\Omega (M,\Delta)
&=&\frac{M^2+\Delta^2}{4G}-\frac 12\sum_{i=1}^{8}\int\frac{d^3p}{(2\pi)^3}|\mu-\eta_{i}|.\label{26}
\end{eqnarray}
Each root $\eta_i$ of the polynomials $P_{\pm}(\eta)$ can be found analytically in the form of a rather cumbersome expression, the procedure is outlined in the Appendix A. In our opinion, there is no mixed phase in the massless NJL model (1). It means that at arbitrary fixed values of chemical potentials the global minimum point (GMP) of the TDP (\ref{07}) lies either on the $M$ axis or on the $\Delta$ axis. This circumstance significantly simplifies the investigation of the phase diagram of the model, since in this case it is enough to study only the projections $F_1(M)\equiv\Omega (M,\Delta=0)$ and $F_2(\Delta)\equiv\Omega(M=0,\Delta)$ of the TDP (\ref{07}) on the $M$ and $\Delta$ axes, correspondingly. \footnote{However, in the particular case when $\mu_5=0$ we actually managed to show that the mixed phase is absent in the massless NJL model (1) (see in Ref. \cite{kkz}). By analogy, one can show that at $\mu_5\ne 0$ it is also absent.} Hence, let us find the functions (projections) $F_1(M)$ and $F_2(\Delta)$.

Looking at the relations (\ref{18}) and (\ref{20}), one can obtain the roots $\eta^{M}_{i}$ of the polynomials $P_\pm(\eta)$ at $\Delta=0$, where
\begin{eqnarray}
\eta^{M}_{1,2,3,4}=\nu \pm\sqrt{M^2+(|\vec p|\pm(\mu_5-\nu_{5}))^2},~~~
\eta^{M}_{5,6,7,8}=-\nu \pm\sqrt{M^2+(|\vec p|\pm(\mu_5+\nu_{5}))^2},
\label{130}
\end{eqnarray}
as well as the roots $\eta^{\Delta}_{i}$ of these polynomials at $M=0$,
\begin{eqnarray}
\eta^{\Delta}_{1,2,3,4}=\nu_{5} \pm\sqrt{\Delta^2+(|\vec p|\pm(\mu_5-\nu))^2},~~~~
\eta^{\Delta}_{5,6,7,8}=-\nu_{5} \pm\sqrt{\Delta^2+(|\vec p|\pm(\mu_5+\nu))^2}.
\label{13}
\end{eqnarray}
Now, taking into account the relations (\ref{130}) and (\ref{13}), one can find the following expressions for the projections $F_{1}(M)$ and $F_{2}(\Delta)$ of the TDP (\ref{26}) on the axes $M$ and $\Delta$, correspondingly,
\begin{eqnarray}
F_{1}(M)\equiv \Omega(M,0)
=\frac{M^2}{4G}-\frac{1}{4\pi^2}\sum_{i=1}^{8}\int_{0}^{\Lambda}d|\vec p||\vec p|^2
|\mu-\eta^{M}_{i}|,\label{F1ref}
\end{eqnarray}
\begin{eqnarray}
F_{2}(\Delta)\equiv \Omega(0,\Delta)
=\frac{\Delta^2}{4G}-\frac{1}{4\pi^2}\sum_{i=1}^{8}\int_{0}^{\Lambda}d|\vec p||\vec p|^2
|\mu-\eta^{\Delta}_{i}|.\label{F2ref}
\end{eqnarray}
To obtain Eqs (\ref{F1ref}) and (\ref{F2ref}) we have used in the integral term of Eq. (\ref{26}) the polar coordinate system and then integrated there over polar angles. Moreover, in addition to $G$, in Eqs (\ref{F1ref}) and (\ref{F2ref}) the cutoff parameter $\Lambda$ is introduced.
In the following we will study the behaviour of the global minimum point of the TDP (\ref{26}) vs chemical potentials for a special set of the model parameters,
$$
G=15.03\, GeV^{-2},\,\,\,\,\,\,\,\,\,\,\,\,\,\,\,\,\Lambda=0.65\, GeV.
$$
In this case at zero chemical potentials one gets for constituent quark mass the value $M=301.58\, MeV$. The same parameter set has been used, e.g., in Refs \cite{Buballa:2003qv,eklim}. The integration in Eqs (\ref{F1ref}) and (\ref{F2ref}) can be carried out analytically but the obtained expressions would be rather involved. So it is  easier to use numerical calculations for evaluation of the integrals.

As a result, we see that in order to find the GMP of the whole TDP (\ref{26}) (or (\ref{07})), one should compare the least values of the functions $F_1(M)$ and $F_2(\Delta)$. By this way, it is clear that there can exist no more than three different phases in the model (1). The first one is the symmetric phase, which corresponds to the global minimum point $(M_0,\Delta_0)$ of the TDP (\ref{26}) of the form $(M_0=0,\Delta_0=0)$. In the CSB phase the TDP reaches the least value at the point $(M_0\ne 0,\Delta_0=0)$. Finally, in the charged PC phase the global minimum point lies at the point $(M_0=0,\Delta_0\ne 0)$. (Notice, that in the most general case the coordinates (condensates) $M_0$ and $\Delta_0$ of the global minimum point depend on chemical potentials.)

Since one of the purposes of the present paper is to prove the possibility of the charged PC phenomenon in dense quark matter (at least in the framework of the NJL model (1)), the consideration of the physical quantity $n_{q}$, called quark number density, is now in order.
This quantity is a very important characteristic of the ground state. It is related to the baryon number density as $n_{q}=3n_B$ because $\mu=\mu_B/3$. Let us present here the ways how expressions for n$_{q}$ can be found in different phases. Recall that in the general case this quantity is defined by the relation
\begin{eqnarray}
n_q=-\frac{\partial\Omega(M_0,\Delta_0)}{\partial\mu}, \label{37}
\end{eqnarray}
where $M_0$ and $\Delta_0$ are coordinates of the GMP of a thermodynamic potential. So in the chiral symmetry breaking phase we have
\begin{align}
n_q(\mu,\mu_5,\nu,\nu_{5})\bigg |_{CSB}=-\frac{\partial\Omega
(M_0,\Delta_0=0)}{\partial\mu}=-\frac{\partial
F_1(M_0)}{\partial\mu}.
\label{38}
\end{align}
Taking into account (\ref{F1ref}) it is not very difficult to get the following expression
\begin{eqnarray}
n_{q}(\mu,\mu_5,\nu,\nu_{5})\bigg |_{CSB}
=\frac{1}{4\pi^2}\sum_{i=1}^{8}\int_{0}^{\Lambda}d|\vec p||\vec p|^2\left\{2\theta(\mu-\eta_{i}^{M_0})-1\right\},
\label{33}
\end{eqnarray}
where $\eta_{i}^{M_0}$ is given by Eq. (\ref{130}) at $M=M_0$.

In a similar way, the particle density in the charged pion condensation phase looks like
\begin{align}
n_q(\mu,\mu_5,\nu,\nu_{5})\bigg |_{PC}=-\frac{\partial\Omega
(M_0=0,\Delta_0)}{\partial\mu}=-\frac{\partial
F_2(\Delta_0)}{\partial\mu}.
\end{align}
Since the quantity $F_2(\Delta_0)$ is defined by Eq. (\ref{F2ref}) at $\Delta=\Delta_0$, one can get
\begin{eqnarray}
n_{q}(\mu,\mu_5,\nu,\nu_{5})\bigg |_{PC}
=\frac{1}{4\pi^2}\sum_{i=1}^{8}\int_{0}^{\Lambda}d|\vec p||\vec p|^2\left\{2\theta(\mu-\eta_{i}^{\Delta_0})-1\right\},
\label{35}
\end{eqnarray}
where $\eta_{i}^{\Delta_0}$ is defined by Eq. (\ref{13}) at $\Delta=\Delta_0$.

Finally, few remarks on the duality properties of the projections $F_1(M)$ and $F_2(\Delta)$. It is clear from Eq. (\ref{19}) as well as from Eq. (\ref{130}) that $F_1(M)$ is invariant under the transformation ${\cal D}_{M}\!\!:\nu_{5}\leftrightarrow \mu_{5}$. So, if dynamically (or for some other reasons) pion condensation is suppressed in the system (i.e. $\Delta=0$), then in the $(\mu_{5},\nu_{5})$-phase diagram the region with CSB phase is arranged symmetrically with respect to the line $\nu_{5}=\mu_{5}$. It means that when we can ignore the appearance of the charged PC phenomenon (e.g. when isospin asymmetry is absent and $\nu=0$), the influence of both $\nu_{5}$ and $\mu_{5}$ on the system is identical.

In a similar way, one can see from Eqs (\ref{21}) and (\ref{13}) that the function $F_2(\Delta)$ is symmetric under the transformation ${\cal D}_{\Delta}\!\!:\nu\leftrightarrow \mu_{5}$. Therefore, if due to some reasons the generation of the chiral condensate is suppressed in the massless NJL model (1) (for example, at low values of the chemical potential $\mu$, etc), then isospin imbalance ($\nu\ne 0$) influences 
the system in the same manner as the chiral imbalance ($\mu_5\ne 0$).

\section{Phase diagram}

\subsection{Dual symmetries of the general $(\mu,\mu_{5},\nu,\nu_{5})$-phase diagram}\label{IVA}

In order to get phase structure of the model one has to find GMP $(M_0,\Delta_0)$ of the thermodynamic potential (\ref{07}). By analogy with the case of $\mu_{5}=0$ (see Ref. \cite{kkz}), where it was shown that in the massless model (1) there is no mixed phase, which corresponds to both $M_0\ne 0$ and $\Delta_0\ne 0$, we will assume that in the general case with  $\mu_{5}\ne 0$ the same is also true. So to study the phase diagram of the massless model (1) one can use the projections $F_1(M)$ (\ref{F1ref}) and $F_2(\Delta)$ (\ref{F2ref}) of this TDP to the axes $M$ and $\Delta$, respectively. It is necessary to determine the GMPs of these projections with respect to $M$ and $\Delta$. Then, one should compare the minimum values of these functions, the result is the GMP $(M_0,\Delta_0)$ of the whole TDP (\ref{07}). (Note, that at least one of the coordinates, $M_0$ or $\Delta_0$, of the obtained GMP is equal to zero.) After this, using numerical calculations, it is necessary to study the behavior of the TDP global minimum point $(M_0,\Delta_0)$ vs chemical potentials. The result is the most general $(\mu,\mu_{5},\nu,\nu_{5})$-phase portrait of the model, i.e. the one-to-one correspondence between any point $(\mu,\mu_{5},\nu,\nu_{5})$ of the four-dimensional space of chemical potentials and possible model phases (CSB, charged PC and symmetric phase). However, it is clear that this four-dimensional  phase portrait is quite bulky and it is rather hard to imagine it as a whole. So in order to obtain a more deep understanding of the phase diagram as well as to get a greater visibility of it, it is very convenient to consider different low-dimensional cross-sections of this general $(\mu,\mu_{5},\nu,\nu_{5})$-phase portrait, defined by the constraints of the form $\nu= const$ or $\mu_5=const$ and $\nu_5=const$, etc. In the next subsections these different cross-sections of the most general phase portrait will be presented. But before that, let us discuss the role and influence both of the main duality ${\cal D}$ (\ref{16}) and constrained dualities ${\cal D_M}$ (\ref{19}) and ${\cal D}_\Delta$ (\ref{21}) of the model on the shape of its different phase portraits.

Let us discuss the form of the most general $(\mu,\mu_{5},\nu,\nu_{5})$-phase portrait of the model.
Suppose that at some fixed particular values of the chemical potentials $\mu$, $\mu_{5}=A$, $\nu=B$ and $\nu_{5}=C$ the global minimum of the TDP (\ref{07}) lies at the point, e.g., $(M=M_0\ne 0,\Delta=0)$. It means that for such fixed values of the chemical potentials the CSB phase is realized in the model. Then it follows from the invariance of the TDP with respect to the main duality transformation ${\cal D}$ (\ref{16}) that at permuted chemical potential values (i.e. at $\nu=C$ and $\nu_{5}=B$ and intact values of $\mu$ and $\mu_{5}=A$) the global minimum of the TDP $\Omega(M,\Delta)$ is arranged at the point $(M=0,\Delta=M_0)$, which corresponds to the charged PC phase (and vice versa). This is the so-called main duality correspondence  in the framework of the model under consideration (or the duality between CSB and charged PC phases). Hence, the knowledge of a phase of the model (1) at some fixed values of external free model parameters $\mu,\nu,\nu_{5},\mu_5$ is sufficient to understand what a phase (we call it a dually conjugated) is realized at rearranged values of external parameters, $\nu\leftrightarrow\nu_{5}$, at fixed $\mu$ and $\mu_{5}=A$. Moreover, different physical parameters such as condensates, densities, etc, which characterize both the initial phase and the dually conjugated one, are connected by the main duality transformation ${\cal D}$. For example, the chiral condensate of the initial CSB phase at some fixed $\mu,\nu,\nu_{5},\mu_5$ is equal to the charged-pion condensate of the dually conjugated charged PC phase, in which one should perform the replacement $\nu\leftrightarrow\nu_{5}$. Knowing the particle density $n_q(\nu,\nu_{5})$ of the initial CSB phase as a function of chemical potentials $\nu,\nu_{5}$, one can find the particle density in the dually conjugated charged PC phase by interchanging $\nu$ and $\nu_{5}$ in the expression $n_q(\nu,\nu_{5})$, etc.

Moreover, one can apply to the point $(\mu,\mu_{5}=A,\nu=B,\nu_{5}=C)$ of the general $(\mu,\mu_{5},\nu,\nu_{5})$-phase portrait the constrained duality invariance ${\cal D}_M$ (\ref{19}) of the TDP, if in this point the CSB phase with chiral condensate $M_0\ne 0$ is arranged. As a result we can conclude that in the point $(\mu,\mu_{5}=C,\nu=B,\nu_{5}=A)$ ($\mu_5$ and $\nu_5$ values are permuted) at least the metastable CSB state can be observed.  However, if due to some reasons the pion condensation is suppressed (e.g., at $\nu=0$), then a stable CSB phase is realized in this point with the same value $M_0$ of the chiral condensate. 
In contrast, if in the initial point $(\mu,\mu_{5}=A,\nu=B,\nu_{5}=C)$ we have the charged PC phase, then (due to the constrained duality invariance ${\cal D}_\Delta$ (\ref{21}) of the model TDP) this phase is at least a metastable one in the point $(\mu,\mu_{5}=B,\nu=A,\nu_{5}=C)$ ($\mu_5$ and $\nu$ values are permuted) with the same pion condensate value, etc.

The main duality transformation ${\cal D}$ of the TDP can also be applied to an arbitrary phase portrait of the model. In particular, it is clear that if we have a most general $(\mu,\mu_5,\nu,\nu_{5})$-phase portrait, then under the duality transformation ${\cal D}$ (which is now understood as a renaming both of the diagram axes, i.e. $\nu\leftrightarrow\nu_{5}$, and phases, i.e. CSB$\leftrightarrow$charged PC) this phase portrait is mapped to itself, i.e. the most general $(\mu,\mu_5,\nu,\nu_{5})$-phase portrait is self-dual. Furthermore, the self-duality of the general $(\mu,\mu_5,\nu,\nu_{5})$-phase portrait means that, e.g., in the two-dimensional $(\nu,\nu_{5})$-phase diagram regions of the CSB and charged PC phases are arranged mirror-symmetrically with respect to the line $\nu=\nu_{5}$.

Another example is the action of the duality transformation $\cal D$ (\ref{16}) on the $(\nu,\mu_5)$-phase diagram at arbitrary fixed values of $\nu_5=A$ and $\mu$. It is clear that under this mapping we obtain the $(\nu_5,\mu_5)$-phase diagram (the axis $\nu$ on the original diagram is replaced by $\nu_5$ on the dually conjugated one) at fixed values of $\nu=A$ and $\mu$ on which two phases, PC and CSB, are rearranged, etc.

\subsection{The case of zero chiral isospin $\mu_{I5}\equiv 2\nu_5$ chemical potential}\label{IVB}

Recall that in the absence of chiral isospin asymmetry, i.e. at $\nu_5=0$, influence of the chiral chemical potential $\mu_5$ on the properties of dense and isospin symmetric ($\mu_I=0$) quark matter was investigated, e.g., in Refs
\cite{andrianov,braguta,braguta2,AndrianovEspriu1,Farias:2016let,Ruggieri:2011xc,Ruggieri:2011wd,Frasca:2016rsi}. In particular, it was shown there that $\mu_5$ is able to catalyze the chiral symmetry breaking phenomenon. In the present subsection we are going to study a more realistic situation, when in addition to $\mu$ and $\mu_5$ the isospin $\mu_I\equiv 2\nu$ chemical potential is also taken into account (but $\nu_5$ is still a zero quantity).

\subsubsection{$(\nu,\mu_5)$-phase diagrams at $\nu_5=0$}\label{IVB1}

\begin{figure}
\includegraphics[width=0.45\textwidth]{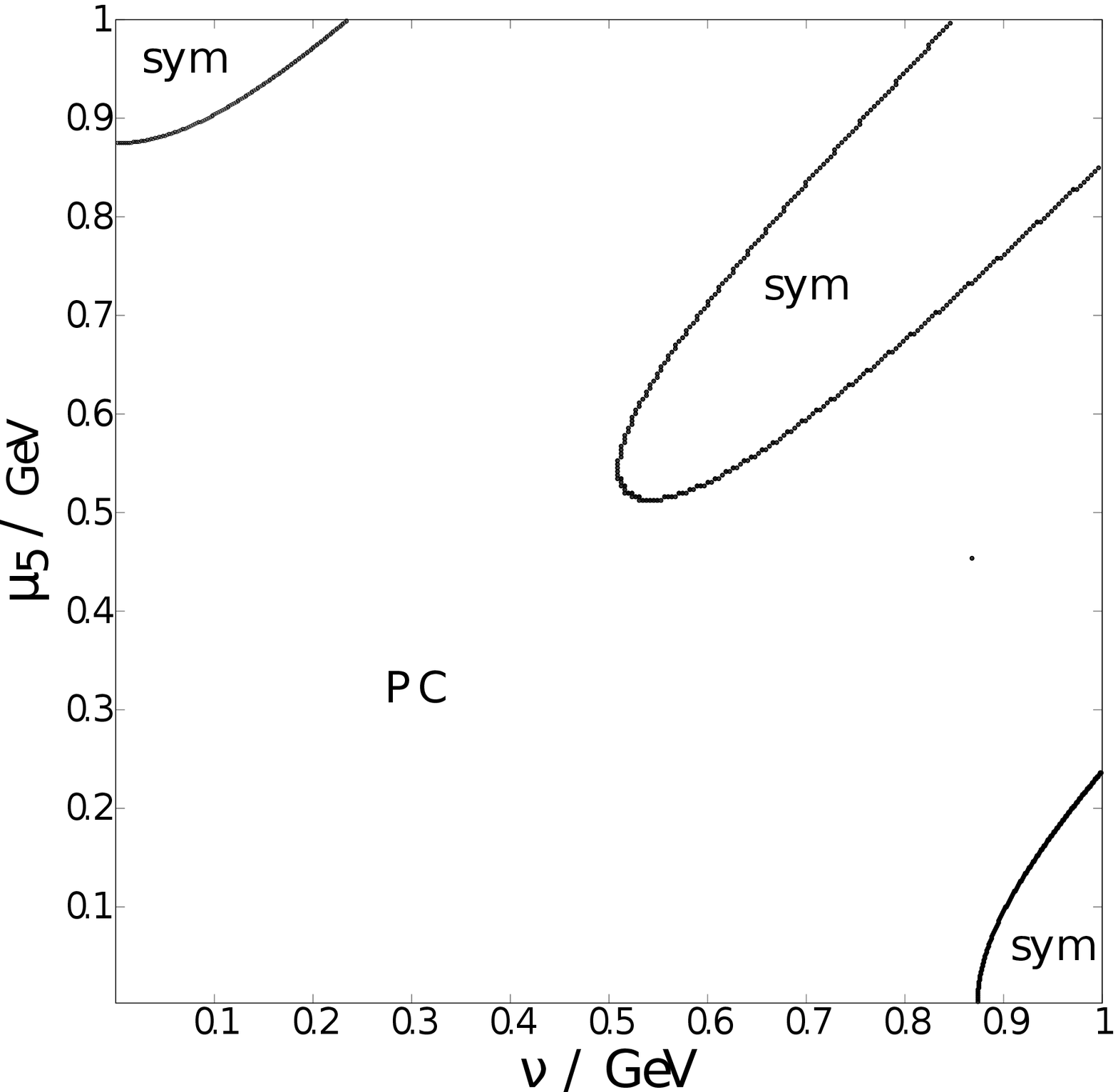}
 \hfill
\includegraphics[width=0.45\textwidth]{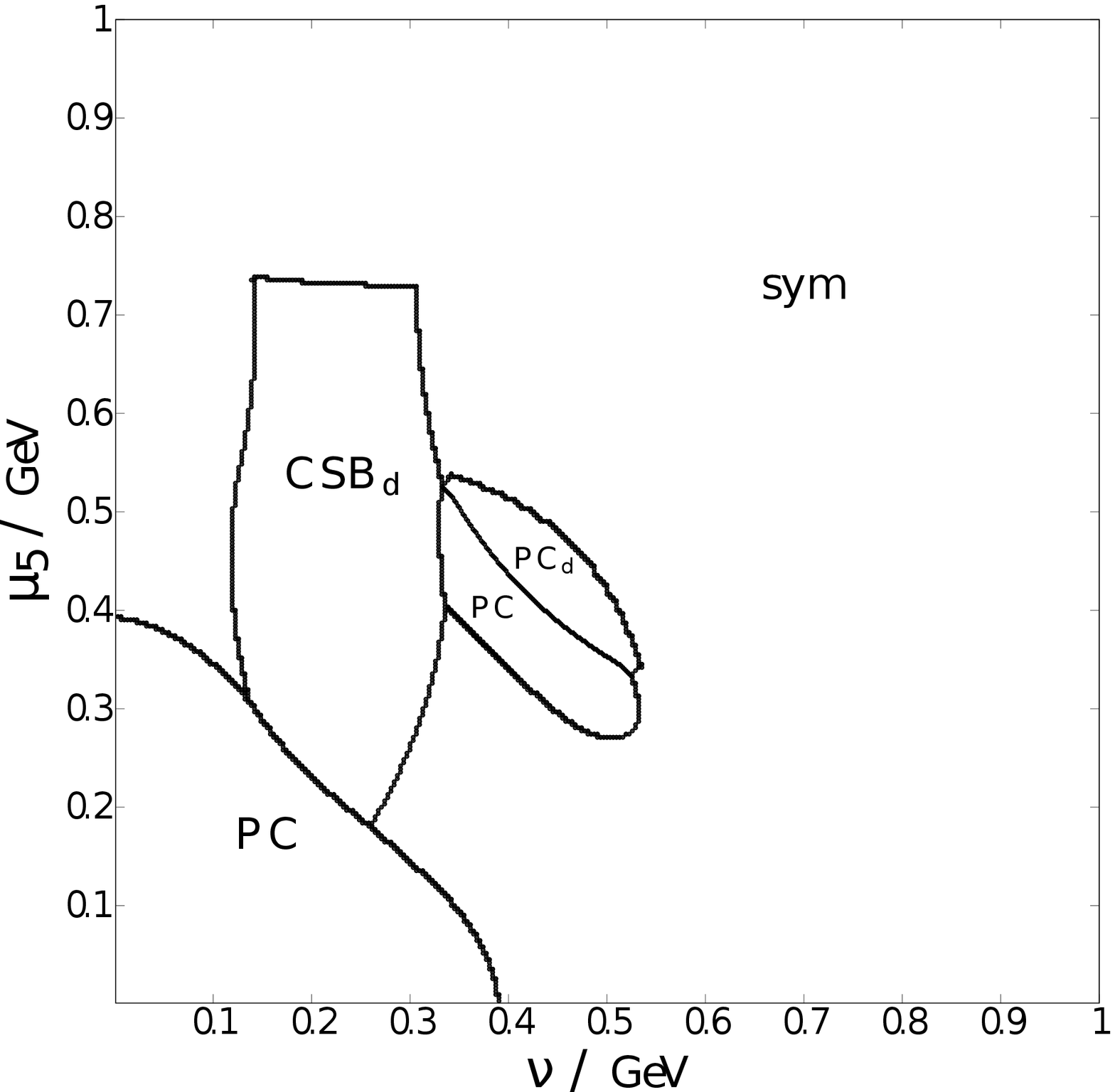}\\
\parbox[t]{0.45\textwidth}{
\caption{ $(\nu,\mu_{5})$-phase diagram at $\mu=0.01$ GeV and $\nu_5=0$. Here PC denotes
the charged pion condensation phase with zero quark number density, ``sym`` is the symmetric phase, where all symmetries are restored.}
 }\hfill
\parbox[t]{0.45\textwidth}{
\caption{
$(\nu,\mu_{5})$-phase diagram at $\mu=0.23$ GeV and $\nu_5=0$. Here CSB denotes the chiral symmetry breaking phase with zero quark number density. CSB$_d$ and PC$_d$ denote the chiral symmetry breaking and charged pion condensation phases with nonzero quark number densities, respectively. Other  notations are presented in Fig. 2. } }
\end{figure}

In our previous paper \cite{kkz} we have investigated the properties of the massless NJL model (1) under the influence of only chiral isospin imbalance, whereas the chiral imbalance effect was ignored, i.e. the case $\mu_5=0$ (and nonzero values of other chemical potentials $\nu_5,\nu,\mu$) was considered there. In this section we are going to consider another limiting and a more physical case when already chiral isospin imbalance is absent, i.e. we study the $\nu_5=0$ cross-sections of the most general $(\mu,\mu_{5},\nu,\nu_5)$-phase portrait of the model (1). As a result, in figures below different $(\nu,\mu_5)$-phase diagrams of the model at different fixed values of $\mu$ are depicted at $\nu_5=0$. Let us discuss what changes in the phase portrait, when instead of $\mu_{I5}$ the chemical potential $\mu_{5}$ acts on the system.

One can see that at values of $\mu\le 0.15$ GeV (see, e.g., Fig. 2) there is no CSB phase. In this case the phase portrait is self-dual with respect to $\nu\leftrightarrow \mu_{5}$ transformation and the charged PC phase (in which quark number density $n_q$ is zero) is arranged symmetrically with respect to the line $\nu=\mu_5$ as it should be according to the duality relation ${\cal D}_{\Delta}$ (\ref{21}). Here we do not present the phase diagram at $\mu=0$ GeV, since it is very similar to the one of Fig. 2 with $\mu=0.01$ GeV, but the region of symmetric phase would be thinner.

Then, as it is seen from Fig. 3, at greater values of $\mu$ the CSB phase appears on the $(\nu,\mu_5)$-phase diagrams (dual symmetry ${\cal D}_{\Delta}$ of the TDP does not prohibit the appearance of the CSB phase). So the charged PC phase is not symmetric  with respect to $\nu\leftrightarrow \mu_{5}$ transformation anymore and the phase diagram is not ${\cal D}_{\Delta}$ self-dual as a whole, but anyway charged PC phase that is present there up to values of  $\nu=0.3$ GeV is symmetric with respect to $\nu\leftrightarrow \mu_{5}$ reflection. Moreover, in Fig. 3 there are some regions with charged PC$_{d}$ phase, in which quark number density $n_q$ is not zero. In contrast to charged PC phase with $n_q=0$, the regions of the PC$_d$ phase are ${\cal D}_{\Delta}$ self-dual there, i.e. they are invariant with respect to ${\cal D}_{\Delta}$ (\ref{21}) transformation. The PC$_{d}$ regions in these figures correspond to rather wide intervals of $\mu_5$ and $\nu$ (from 0.35 Gev to 0.55 Gev). Hence, at nonzero $\nu$ the generation of the charged PC$_d$ phase is possible in the system even at $\mu_5\ne 0$, but only for chemical potential $\mu$ values from a rather narrow interval of not so high $\mu\in (0.21\div 0.25)$ GeV (only at comparatively low values of $\mu$, meaning at not so high baryon densities). Our investigations of the case $\nu_{5}=0$ show that  for values of $\mu$ outside this interval there is no any generation of the charged pion condensation phase with nonzero particle density $n_q$ by chiral chemical potential $\mu_{5}$.

Moreover, starting from $\mu=0.35$ GeV different $(\nu,\mu_5)$-phase portraits do not contain charged PC phase at all (even with $n_q=0$). One can also see that in this case, i.e. at $\mu>0.35$ GeV, the CSB phase, namely its shape and position in the $(\nu,\mu_5)$-phase diagram as well as its behavior vs. $\mu$, resembles (compare Figs 4 and 5) or even equal (where there is no PC phase) to the CSB phase in the $(\nu,\nu_5)$-phase diagrams of the model (1) at $\mu_5=0$ \cite{kkz}. For example, in Fig. 4 we have depicted the $(\nu,\mu_5)$-phase portrait at $\mu=0.4$ GeV and $\nu_5=0$. Comparing it with Fig. 5, where the $(\nu,\nu_5)$-phase portrait at $\mu_5=0$ and at the same value of $\mu=0.4$ GeV is depicted (see also Fig. 6 in Ref. \cite{kkz}), we see that in both diagrams the CSB phase takes the shape of a sole of a boot that points at the $\nu$-axis (at the value of $\nu=\mu $) and, except a small region, their sizes, positions and forms are equal. Such a coincidence can be explained by the constrained duality ${\cal D}_{M}$ (\ref{19}) of the model.
It tells us that if at the point $(\mu,\mu_5=0,\nu,\nu_5=A)$ the CSB phase is arranged, then in the (dually ${\cal D}_{M}$ conjugated) point $(\mu,\mu_5=A,\nu,\nu_5=0)$ the CSB phase must also be realized, if charged PC phenomenon is suppressed in the system. And just this constraint is valid for the phase diagram of Fig. 4, where PC phase is absent. So, if in the point $(\nu,\nu_5=A)$ of the $(\nu,\nu_5)$-phase diagram of Fig. 5 we have CSB phase, then it is also arranged in the point $(\nu,\mu_5=A)$ of the $(\nu,\mu_5)$-phase portrait of Fig. 4.
(the reverse is not necessarily true because the PC phase is not suppressed in Fig. 5 everywhere). Hence, knowing the position of the CSB phase in the $(\nu,\nu_5)$-phase diagram at $\mu_5=0$, we can predict (due to the constrained duality relation ${\cal D}_{M}$ (\ref{19})) the position of this phase in the $(\nu,\mu_5)$-phase portrait at $\nu_5=0$, if there is a restriction that prohibits the existence of the charged PC phase in the system.

Another interesting correspondence between influences of chiral ($\mu_5\ne 0$) and isospin ($\nu\ne 0$) imbalances on the model (1) can be found from the $(\nu,\mu_5)$-phase diagram at $\nu_5=0$ and $\mu=0$ (that looks very similar to the one of Fig. 2 at $\mu=0.01$ GeV as was mentioned above), which does not contain any points with CSB phase. In this case, due to the duality ${\cal D}_{\Delta}$, one can find out that at the points $(\mu=0,\mu_5=A,\nu=0,\nu_5=0)$ and $(\mu=0,\mu_5=0,\nu=A,\nu_5=0)$ of the general phase diagram the charged PC phase is arranged with the same value of the condensate $\Delta_0$, which is a GMP of the function $F_2(\Delta)$ (\ref{F2ref}) at $\nu=\nu_5=0$. However, in this case the function $F_1(M)$ (\ref{F1ref}) with GMP $M_0$ is equal to $F_2(\Delta)$ (as it follows from (\ref{130}) and (\ref{13})), so $M_0=\Delta_0$, and the minima $(M_{0},0)$ and $(0,\Delta_{0})$ of the TDP (\ref{07}) are degenerate. In the case of $\nu=0$ it makes sense to choose $(M_{0},0)$ minimum because there should not be charged PC in this case.
So one can conclude that at any value of $A$ the quark condensate $M_{0}$ at $\mu_5=A$, $\nu=0$ exactly equals to charged pion condensate $\Delta_{0}$ at $\nu=A$, $\mu_5=0$. (This conclusion can be made also from the ($\nu,\nu_{5}$)-phase diagram at $\mu=0$ GeV from \cite{kkz} with the use of duality ${\cal D}_{M}$.)

Earlier, it was established by lattice simulations and by effective model calculations that $\mu_{I}$ generates charged pion condensation \cite{Brandt:2016zdy,Son:2000xc} and $\mu_{5}$ generates chiral symmetry breaking \cite{braguta2} (catalysis of dynamical chiral symmetry breaking by $\mu_{5}$). Our investigations show that by the duality ${\cal D}_{\Delta}$ (or by the duality ${\cal D}_{M}$) these phenomena should look exactly the same.
And the ($\nu,T$) phase diagram of the model should coincide with ($\mu_{5},T$) phase diagram with the change PC$\leftrightarrow$CSB, etc.
Of course, all these arguments holds only in the chiral limit and in the leading order of the large-$N_c$ approximation, but since zero current quark mass seems to be a good approximation it could hold approximately in reality.
\begin{figure}
\includegraphics[width=0.45\textwidth]{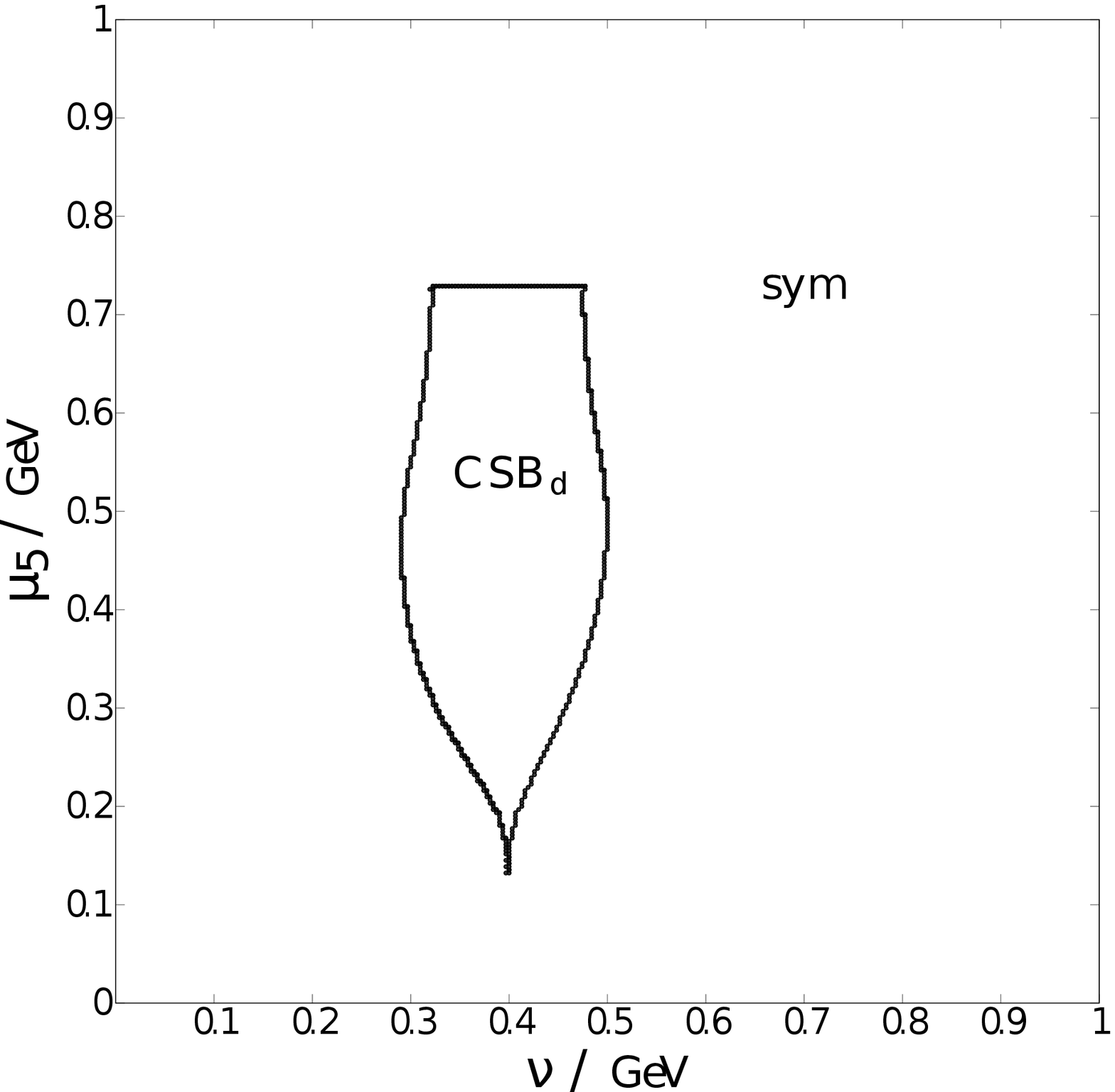}
 \hfill
\includegraphics[width=0.45\textwidth]{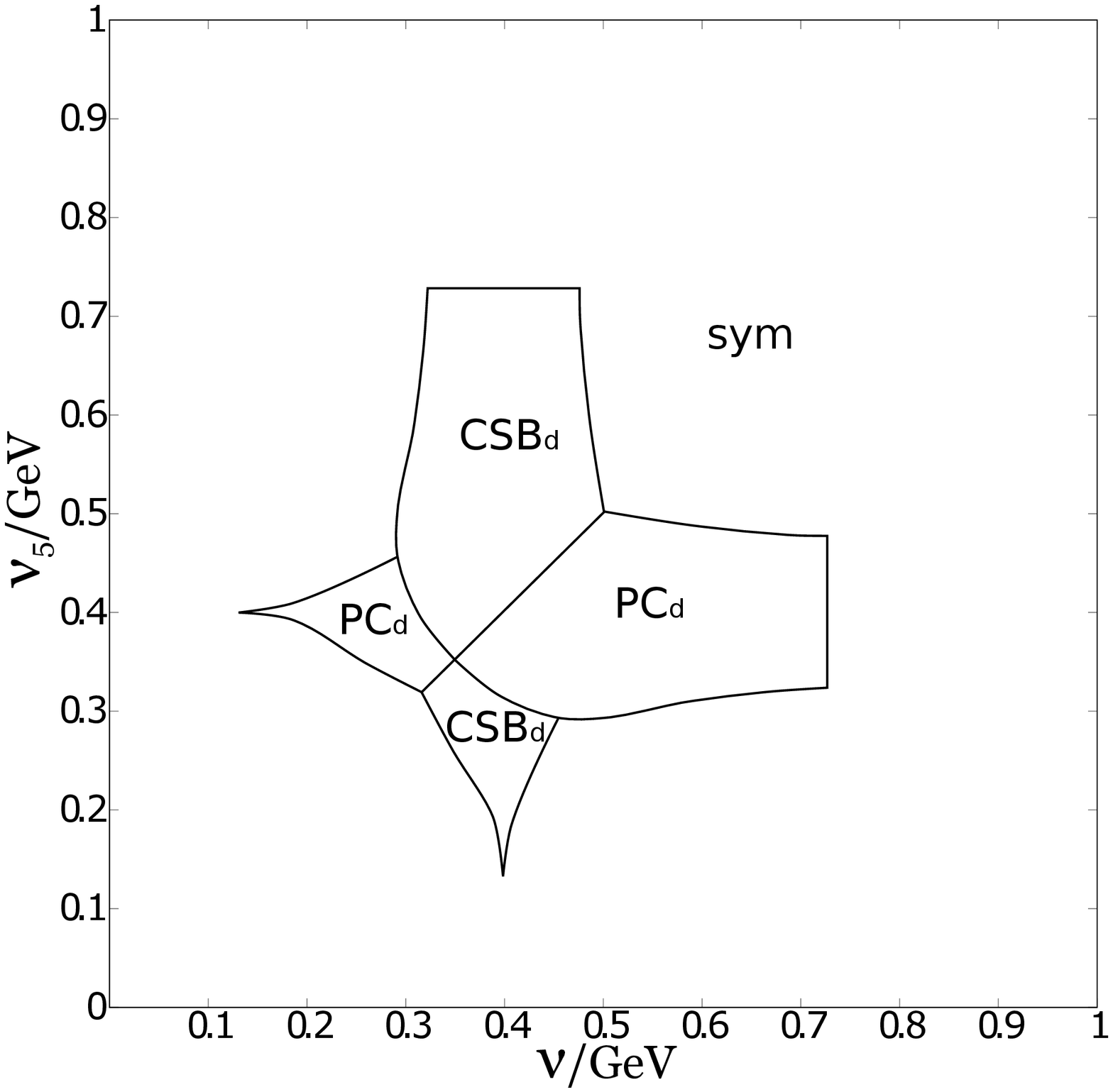}\\
\parbox[t]{0.45\textwidth}{
\caption{ $(\nu,\mu_{5})$-phase diagram at $\mu=0.4$ GeV and $\nu_5=0$. All designations as in previous Figs 2,3.}
 }\hfill
\parbox[t]{0.45\textwidth}{
\caption{ $(\nu,\nu_{5})$-phase diagram at $\mu=0.4$ GeV and $\mu_5=0$. All designations as in previous Figs 2,3.} }
\end{figure}

\subsubsection{$(\mu_5,\mu)$-phase diagrams at $\nu_5=0$}\label{IVB2}

Earlier, it was shown in Ref. \cite{braguta2} that even in the case of unphysical case of weak coupling (when there is no chiral symmetry breaking) and at $\mu=0$ arbitrary small chiral chemical potential $\mu_5$ induces (catalyses) CSB phase in the NJL model at $\nu=0$ and $\nu_5=0$. But in this paper one flavour NJL model was considered (there is obviously no PC phenomenon there)
So it was concluded that there is a catalysis of CSB by chiral $\mu_{5}$ chemical potential in the case of one flavour.
However, this property of $\mu_5$ can be trivially generalized to the case of two quark flavours, if one ignores PC phenomenon. Indeed, look at Fig. 6, where $(\mu_5,\mu)$-phase portrait of two-flavored NJL model (1) is presented at zero values of $\nu$ and $\nu_5$.

Since at $\nu=0$, $\mu_5=0$, but $\nu_5\ne 0$ and $\mu\ne 0$ only the CSB  or symmetrical phases can be realized (see corresponding phase portraits in Ref. \cite{kkz}), i.e. the charged condensate $\Delta$ is equal to zero, one can apply to the phase diagram of Fig. 6 the ${\cal D}_{M}$ duality transformation (\ref{19}). In this case we should only rename the axis $\mu_5$ in the diagram of Fig. 6 in favor of $\nu_5$. As a result, we obtain a $(\nu_5,\mu)$ diagram with constraints $\mu_5=0$ and $\nu=0$, in which only the CSB phase is presented. So at $\mu_5=0$ and $\nu=0$ chiral isospin $\mu_{I5}$ chemical potential catalyses CSB in the framework of the massless NJL model (1) in a similar way, as it occurs in the model under consideration by the action of the $\mu_5$ at $\nu_5=0$ and $\nu=0$ (see also in Ref. \cite{braguta2}). Hence, the catalysis of the CSB phenomenon by $\mu_{I5}$ chemical potential at $\nu=0$ and $\mu_5=0$, is dually-${\cal D}_{M}$ conjugated to the catalysis of the CSB by $\mu_{5}$ (at $\nu=0$ and $\nu_5=0$).

Furthermore, in the framework of the model under consideration the chemical potential $\mu_5$ is able to generate (to catalyse) the charged PC phase as well (at $\nu_5=0$ and $\nu=0$). Indeed, applying to a phase diagram of Fig. 6 the dual transformation ${\cal D}$, one can obtain the dually conjugated phase diagram, which is just Fig. 6 but only with renamed phases, CSB$\rightarrow$PC and CSB$_d\rightarrow$PC$_d$.
It means that at $\mu\ne 0$ and $\mu_5\ne 0$, but at $\nu_5=0$ and $\nu=0$, the TDP (\ref{07}) has two degenerated global minima, first of them, i.e. the point of the form $(M_0\ne 0,\Delta=0)$, corresponds to CSB phase (see in Fig. 6), the second -- the point of the form $(M_0=0,\Delta_0\ne 0)$ -- to the charged PC phase (it is presented in the dual mapping of Fig. 6). The degeneracy of these ground states means that for such values of the chemical potentials in the space, filled, e.g., with CSB  phase, a bubble of the charged PC phase (and vice versa) can be created, i.e. one can observe in  space the mixture (or coexistence) of these two phases.

Since at $\nu=0$ and $\nu_5=0$ one has two degenerated global minima,
in order to get catalysis of CSB by $\mu_{5}$ one need to choose one of the two degenerated global minima, namely CSB phase. But if we have even the infinitesimally small value of isospin chemical potential $\nu$, then the global minimum corresponding to the PC phase becomes deeper than the CSB counterpart, and there is no any catalysis of CSB by chiral chemical potential $\mu_{5}$. So the catalysis of CSB found in Ref. \cite{braguta2} is valid only if isospin chemical potential $\nu$ exactly equals to zero.
Note that the ability to catalyse both CSB and charged PC phenomena at $\nu=0$ is inherent only to the chiral chemical potential $\mu_5$. The chiral isospin chemical potential $\mu_{I5}$ is not able to generate charged PC phase at $\mu_5=0$ and $\nu=0$ \cite{kkz} and in this case one has only one global minimum corresponding to CSB phase. And even at small values of $\nu$ chiral isospin chemical potential $\mu_{I5}$ is able to catalyse CSB (the only requirement for this is $\nu_{5}>\nu$).  So one can conclude that the catalysis of the CSB by chiral isospin chemical potential  $\mu_{I5}$ is even stronger than the one by chiral chemical potential $\mu_{5}$.

\begin{figure}
\includegraphics[width=0.57\textwidth]{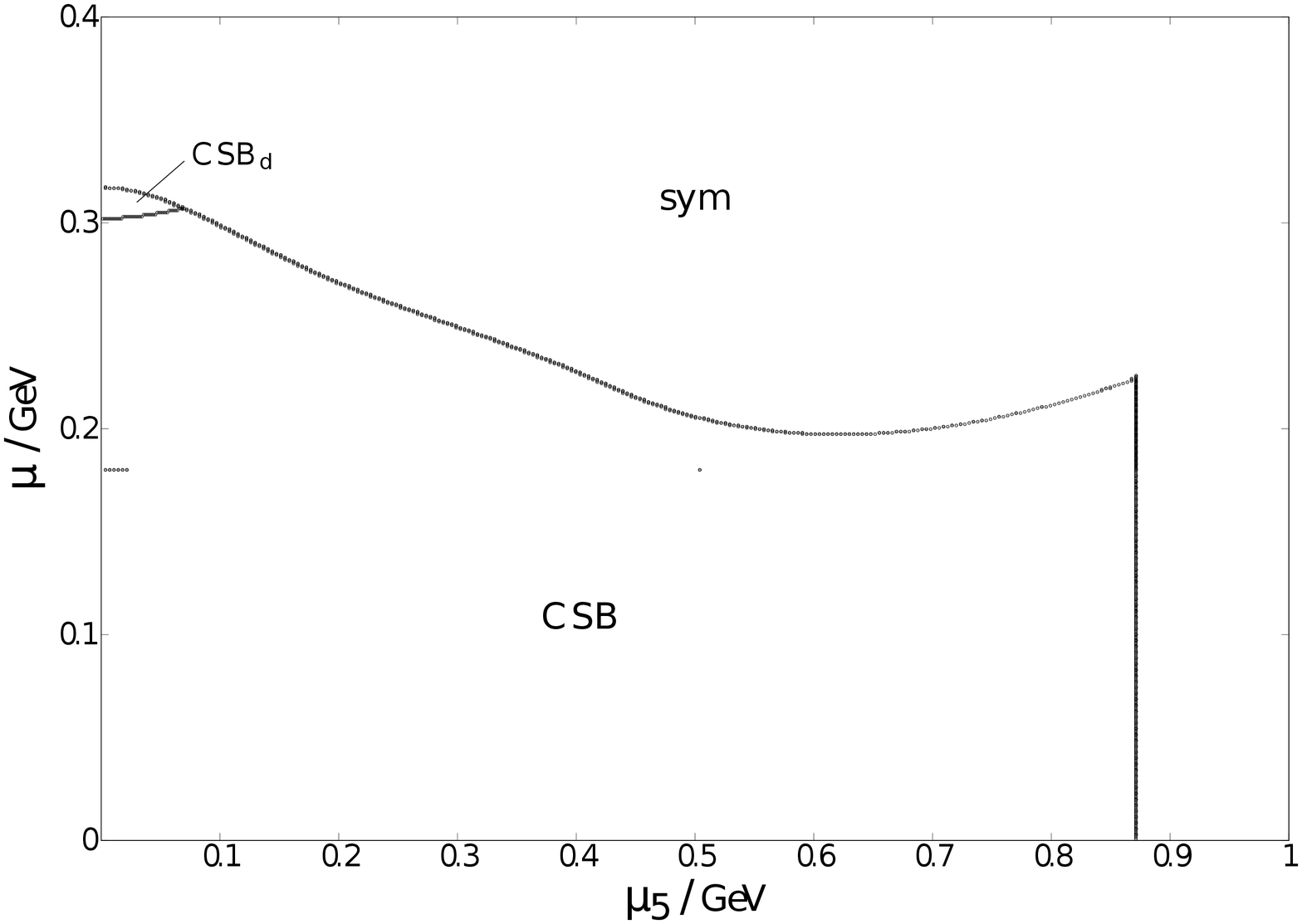}
 \hfill
\includegraphics[width=0.41\textwidth]{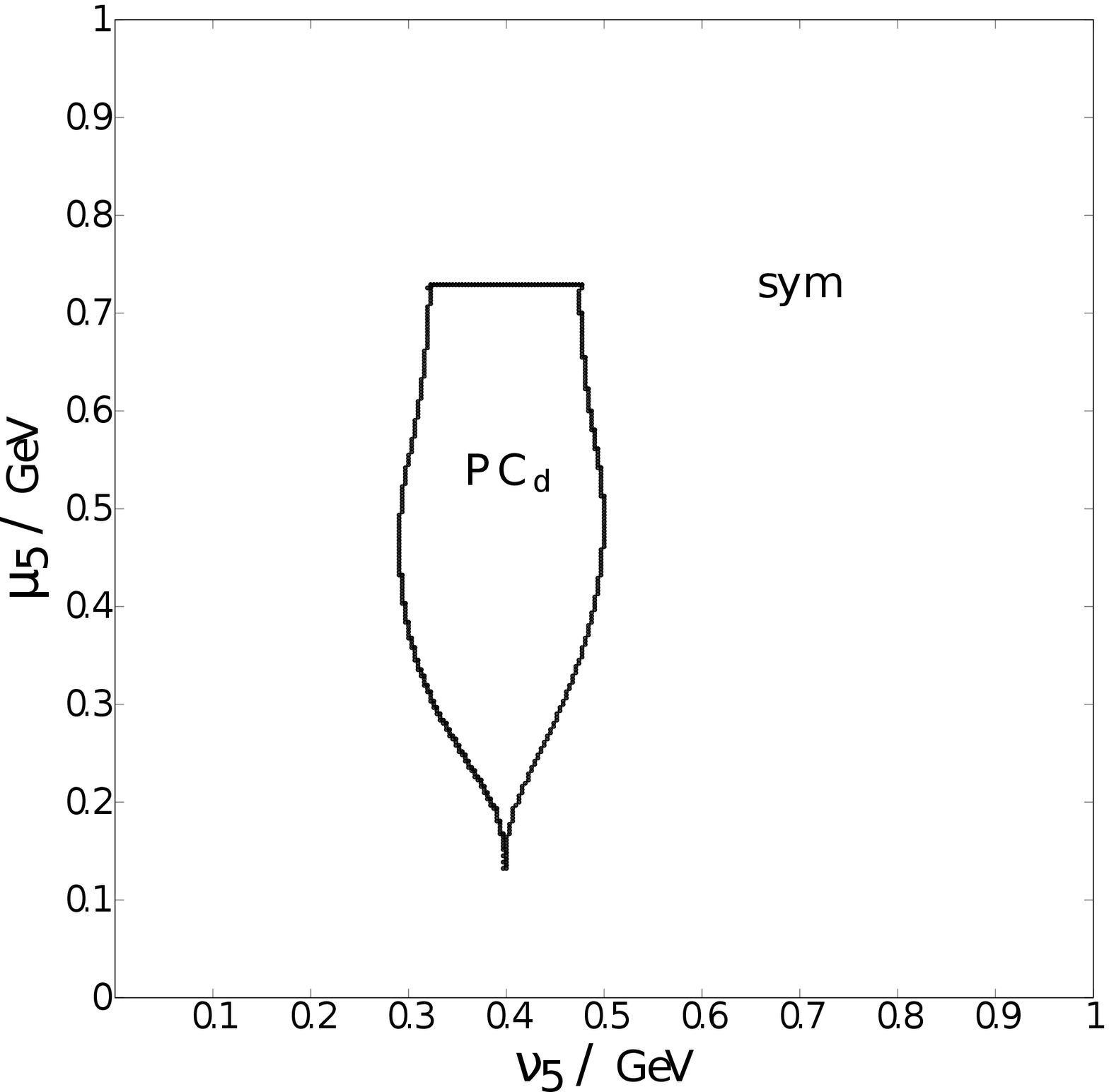}\\
\parbox[t]{0.45\textwidth}{
\caption{  $(\mu_5,\mu)$-phase diagram at $\nu_5=0$ and $\nu=0$. All designations as in previous Figs 2-5.}
 }\hfill
\parbox[t]{0.45\textwidth}{
\caption{ $(\nu_5,\mu_5)$-phase diagram at $\nu=0$ and $\mu=0.4$ GeV. It is a dual $\cal D$ mapping of the phase diagram of Fig. 4.
All designations as in previous Figs 2-5.} }
\end{figure}

\subsection{General case: $\mu_5\ne 0$ and $\nu_5\ne 0$}\label{IVC}

Up to now, i.e. in the section \ref{IVB}, we have investigated the influence of only chiral imbalance ($\mu_5\ne 0$) on quark matter with nonzero both baryon and isospin densities. But the possibility for the chiral isospin asymmetry of the system was ignored, so we have considered there the case $\nu_5=0$. However, since $\mu_5\ne 0$ catalyses CSB \cite{braguta2}, whereas $\nu_5\ne 0$ promotes charged PC \cite{kkz}, it is interesting to study their combined effect on the system. So in the following we are going to consider the case, when two types of chiral asymmetry are present, $\mu_5\ne 0$ and $\nu_5\ne 0$.

\subsubsection{The case of zero isospin $\mu_{I}\equiv 2\nu$ chemical potential}

In this section we consider the case of zero  isospin imbalance ($\mu_I=0$) and will discuss the simultaneous action of the chiral $\mu_5$ chemical potential (it catalyses the CSB phase) and chiral isospin $\nu_5$ chemical potential (as it was shown in \cite{kkz}, it promotes the charged PC phenomenon) on the phase structure of the model.

It turns out that at $\nu=0$ one does not have to calculate anything
because there is a simpler way, which is based on the main duality invariance (\ref{16}) of the TDP. So one can apply the duality transformation $\cal D$ (\ref{16}) both to the $(\nu,\mu_{5})$- and $(\mu_{5},\mu)$-phase diagrams, \footnote{The procedure of applying the main duality transformation $\cal D$ (\ref{16}) to different phase diagrams is presented at the end of section IV A.} obtained in the case of $\nu_{5}=0$ (see previous section \ref{IVB}) in order to find the dually conjugated phase diagrams of the case $\nu=0$. Hence, to find, e.g., the $(\nu_{5},\mu_{5})$-phase diagram at $\mu=0.4$ GeV and $\nu=0$, we should start from the corresponding $(\nu,\mu_{5})$-phase diagram at $\mu=0.4$ GeV and $\nu_{5}=0$ of Fig. 4 and make the simplest replacement of the notations in this figure: $\nu\leftrightarrow\nu_{5}$, PC$_{d}\leftrightarrow$ CSB$_{d}$ (note, the symmetric phase is intact under the dual transformation $\cal D$). The result is the $(\nu_{5},\mu_{5})$-phase portrait of the model at $\mu=0.4$ GeV and $\nu=0$ (see Fig. 7). It is clear from this figure that even at zero value of the isospin chemical potential $\mu_{I}\equiv 2\nu$ there is a possibility for the generation of the charged PC$_{d}$ phase in quark matter at nonzero values of the chiral chemical potential $\mu_5$. But only when $\nu_5\ne 0$.
\begin{figure}
\includegraphics[width=0.45\textwidth]{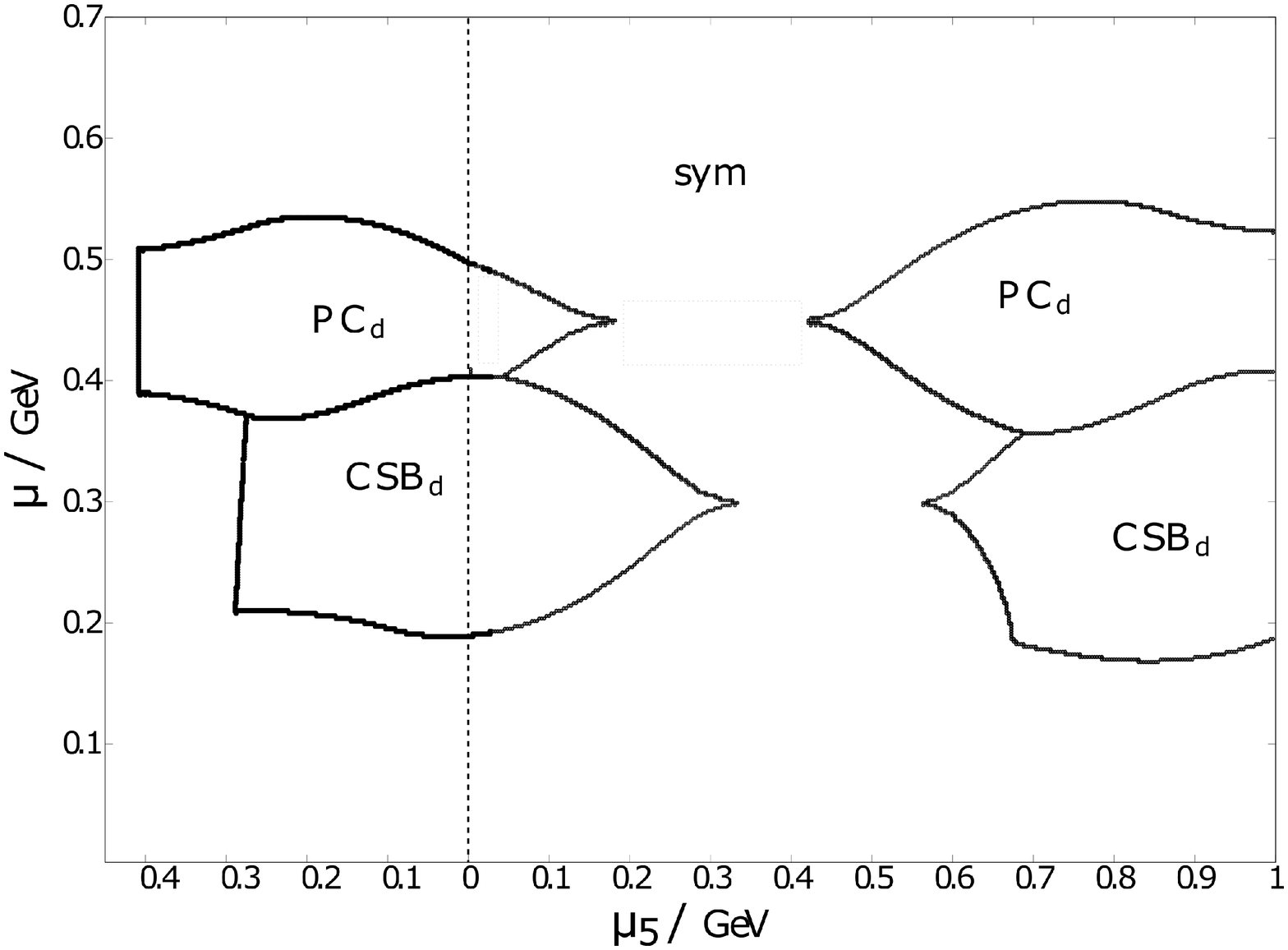}
 \hfill
\includegraphics[width=0.45\textwidth]{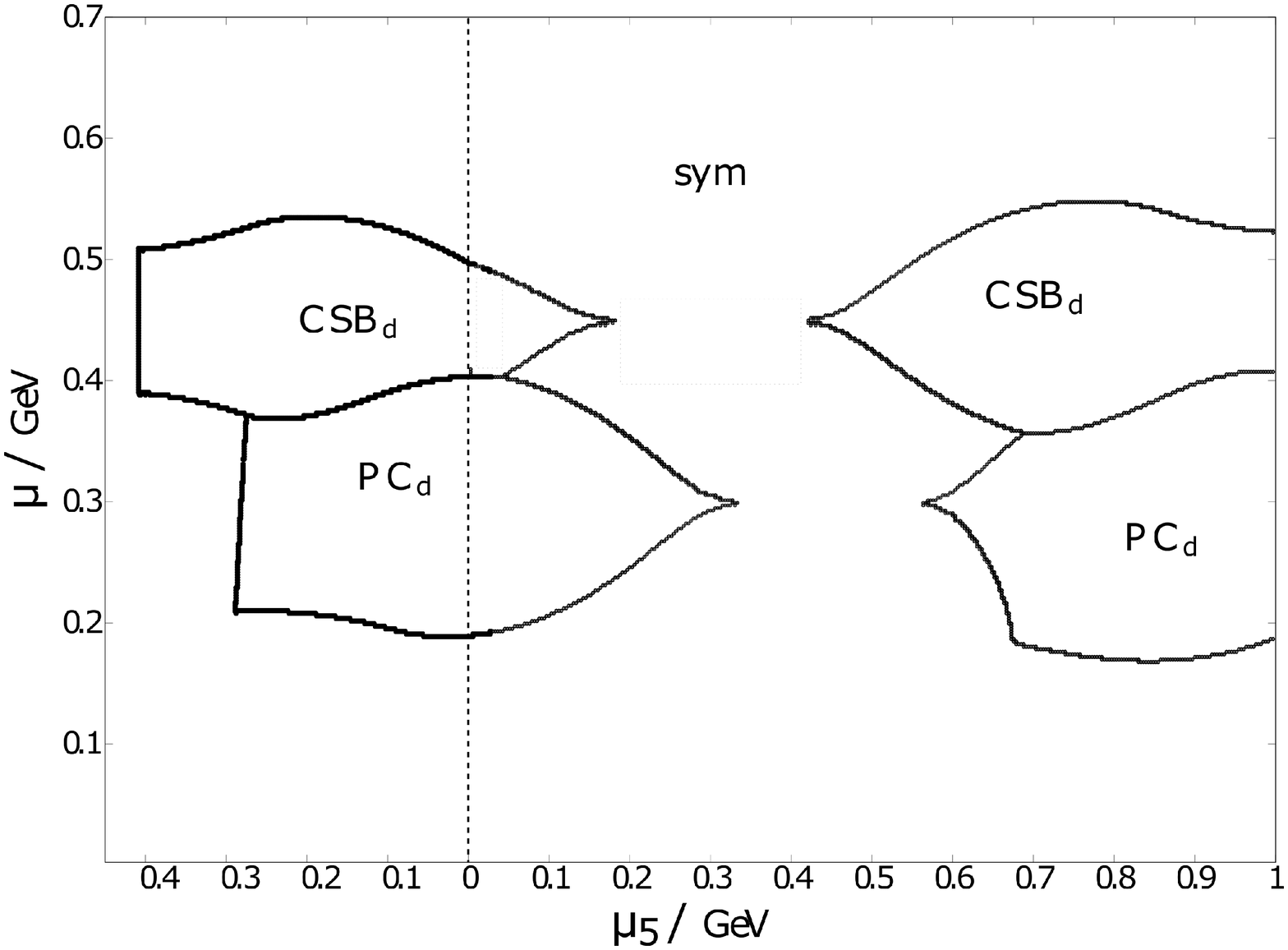}\\
\parbox[t]{0.45\textwidth}{
\caption{ $(\mu_5,\mu)$-phase diagram at $\nu_5=0.45$ GeV and $\nu=0.3$ GeV. It is a dual $\cal D$ mapping of the phase diagram of Fig. 9. All designations as in previous Figs 2-5.}
 }\hfill
\parbox[t]{0.45\textwidth}{
\caption{ $(\mu_5,\mu)$-phase diagram at $\nu_5=0.3$ GeV and $\nu=0.45$ GeV. It is a dual $\cal D$ mapping of the phase diagram of Fig. 8.
All designations as in previous Figs 2-5.} }
\end{figure}

\subsubsection{Other cross-sections of the most general $(\mu,\mu_5,\nu,\nu_{5})$-phase portrait at $\nu\ne 0$ }\label{IVC2}

So far we have considered phase portraits, provided that one of the chemical potentials $\mu_5$, $\nu$ or $\nu_{5}$ is zero. As it is clear from the remark just after  Eq. (\ref{10}), in this case one can consider only nonnegative values of the remaining chemical potentials. However, when all three chemical potentials $\mu_5$, $\nu$ and $\nu_{5}$ are nonzero, we cannot confine ourselves only to positive signs of these quantities in order to establish a complete phase picture of the model. But to simplify the consideration, one can use symmetries (dualities) of the TDP under the reversal of the sign of chemical potentials (see the discussion at the end of Section II). So in this case it is enough to take two (arbitrary) of them as nonnegative quantities and the remaining chemical potential can have an arbitrary sign. Consequently, below several typical phase portraits are presented, where we assume for definiteness that $\nu\ge 0$ and $\nu_{5}\ge 0$, but $-\infty<\mu_5<\infty$. Moreover, these diagrams clearly illustrate the fact that there is a duality between CSB and charged PC in the case of nonzero $\mu_5$ and $\nu_{5}$.

Using above mentioned restrictions on the values of $\mu_5$, $\nu$ and $\nu_{5}$, we have depicted in Figs 8 and 9 the $(\mu_5,\mu)$-phase diagrams of the model. Since the first diagram corresponds to fixed values of $\nu=0.3$ GeV and $\nu_{5}=0.45$ GeV, and the second one is for the same, but interchanged values of these chemical potentials, i.e. for fixed $\nu=0.45$ GeV and $\nu_5=0.3$ GeV, we see that the phase portraits of Figs 8, 9 are dually $\cal D$ symmetric to each other.

Finally, in Figs 10, 11 we have presented two $(\nu,\nu_5)$-phase diagrams at different fixed values of the remaining chemical potentials $\mu$ and $\mu_5$. Each of these diagrams is a self-dual with respect to duality transformation $\cal D$ of an arbitrary phase portrait of the model (see at the end of the section \ref{IVA}). It means that charged PC and CSB phases in a such diagram are arranged mirror-symmetrically to each other with respect to the line $\nu=\nu_5$.

As a result, we see that in the most general case when, in addition to $\mu\ne 0$, other chemical potentials $\mu_5$, $\nu$ and $\nu_{5}$ are also nonzero, the charged PC phase can be generated in dense quark matter.

Let us now try to understand the whole most general $(\mu,\mu_5,\nu,\nu_{5})$-phase portrait of the model. In order to do it, it is easier to consider the most general phase portrait in terms of the  $(\nu,\nu_{5})$-phase portrait at all values of $\mu>0$ and $-\infty<\mu_{5}<\infty$. The behavior of the $(\nu,\nu_{5})$-phase diagram with respect to changing of the value of $\mu$ at $\mu_{5}=0$ has been considered in \cite{kkz}. First, let us recall briefly this phase diagram. From some values of $\mu$ there appear two regions of PC$_d$ and CSB$_{d}$ phases that have the form of soles
of boots 
and with increase of $\mu$ these regions drift sidewise (if you imagine a sole
of a boot) to the higher values of $\nu_{5}$ and  $\nu$ respectively (see Fig. 5).
Now let us consider nonzero values of chiral chemical potential $\mu_5$.  Recall that in order to consider the whole phase diagram one has to consider all values of chemical potential $\mu_5$.  When 
$\mu_{5}$ is greater than zero and increases, then these PC$_d$- and CSB$_{d}$-phase regions goes backwards (heels moves ahead) to the higher values of $\nu$ and $\nu_{5}$ respectively, and from the axes $\nu=0$ and $\nu_{5}=0$ there appear other regions of PC$_d$ and CSB$_{d}$ phases that also look like soles of boots but without heels 
(see Fig. 10). When $\mu_{5}<0$ 
and its absolute value increases then the PC$_d$- and CSB$_{d}$-phase regions slide forwards (toe of the boot ahead) to the smaller values of $\nu$ and $\nu_{5}$, respectively (in the opposite direction to the $\mu_{5}>0$ case). This can be seen from Fig. 11. It is clear that in both of these cases there can be PC$_{d}$ phase at zero values of isospin chemical potential ($\nu=0$).
Let us now mention what changes when one increase baryon chemical potential $\mu$. The regions of PC$_d$ and CSB$_{d}$ phases in that respect behaves in exactly the same way as in the case $\mu_{5}=0$, namely they go sidewise to the higher values of $\nu_{5}$ and  $\nu$ respectively.
\begin{figure}
\includegraphics[width=0.45\textwidth]{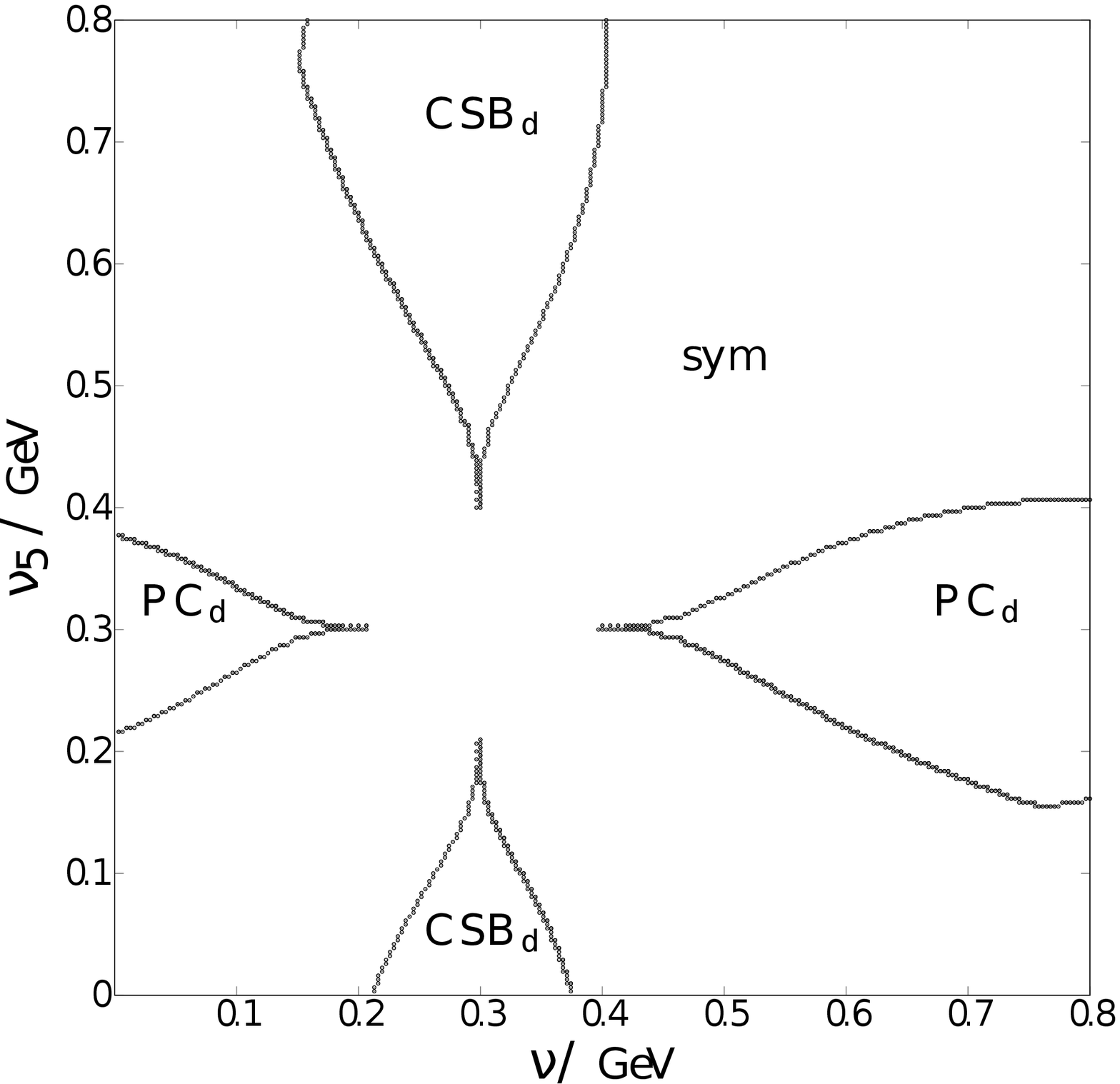}
 \hfill
\includegraphics[width=0.45\textwidth]{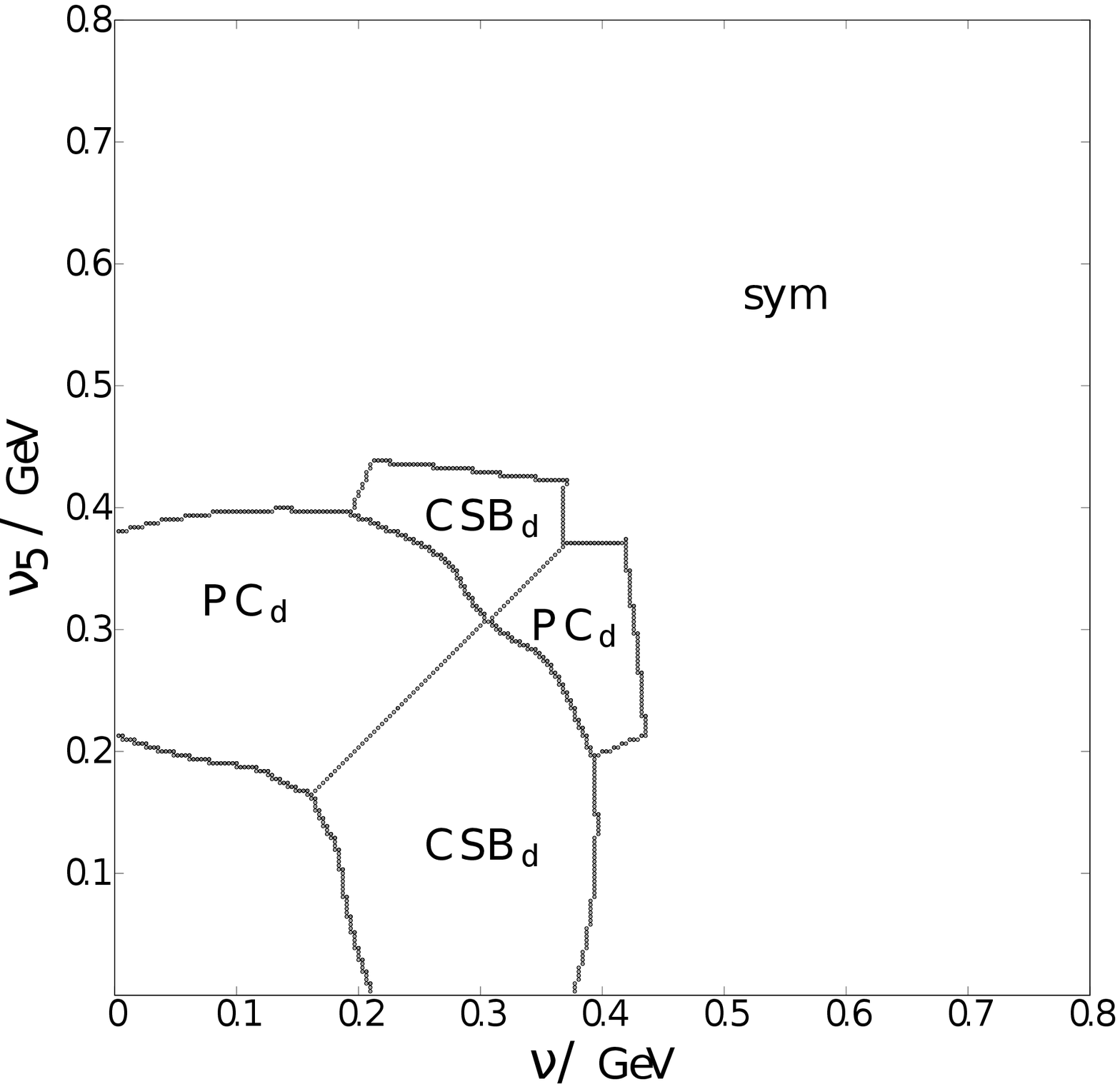}\\
\parbox[t]{0.45\textwidth}{
 \caption{ $(\nu,\nu_{5})$-phase diagram at $\mu_{5}=0.3$ GeV and $\mu=0.3$ GeV. All designations as in previous Figs 2-5.}
 }\hfill
\parbox[t]{0.45\textwidth}{
\caption{ $(\nu,\nu_{5})$-phase diagram at $\mu_{5}=-0.3$ GeV  and $\mu=0.3$ GeV. All designations as in previous Figs 2-5.} }
\end{figure}

\section{Summary and Conclusions}

In this paper the influence of isotopic, chiral and chiral isospin imbalances on phase structure of dense quark matter has been investigated in the framework of the massless (3+1)-dimensional NJL model (1) with two quark flavors in the large-$N_{c}$ limit. It means that we have investigated phase structure of this NJL model at nonzero baryon $\mu_B$, isospin $\mu_I\equiv 2\nu$, chiral isospin $\mu_{I5}\equiv 2\nu_5$ and chiral $\mu_5$ chemical potentials (see Lagrangian (1)).

Earlier, the effect of only $\mu_B,\mu_I,\mu_{I5}$ chemical potentials (i.e. at $\mu_5=0$) on phase structure of quark matter has been considered in the framework of massless NJL$_{2}$ and NJL$_{4}$ models in Refs \cite{ekk,kkz}, where it was shown that $\mu_{I5}$ promotes charged PC phase with {\it nonzero baryon density} (in Refs \cite{ekk,kkz} and in the present consideration this phase is denoted as charged PC$_d$ phase). Moreover, it was established in these papers that in the leading order of the large-$N_c$ approximation there is a duality $\cal D$ between CSB and charged PC phenomena. It is due to a symmetry of the TDP with respect to the transformation (\ref{16}).

Let us say a few words about the choice of using the NJL model.
As it has been said in the Introduction lattice QCD in this case is impossible because of the notorious sign problem, so one has to use effective models in this situation. Chiral symmetry breaking and pion condensation phenomena are quite properly described in the framework of NJL model and it is straightforward to formulate the model in the considered environment.
As it has been also discussed in the Introduction description of the QCD phase structure in terms of PNJL model is a much better approximation and enriches it with a possibility to tackle the confinement/deconfinement regimes. To include two more order parameters (in PNJL model in the case of non-zero baryon chemical potential we need two parameters due to the fact that $\Phi\neq\bar\Phi$, where $\Phi$ is the Polyakov loop parameter) one need much more computation power, although it probably should not change qualitatively the results of our paper, only probably shifts slightly some phase transition lines of CSB, PC and symmetrical phases.
Of course, it can give even better quantitative agreement with QCD phase structure but even this feature requires sometimes additional steps.
For example, the critical temperature in the PNJL model is exaggerated even compared with NJL model one, which is a little bit larger than the QCD simulations predict in the first place. To deal with this feature of the model and to reproduce the results of lattice QCD with imaginary baryon chemical potential one adds the dependence of the coupling on the Polyakov loop and study the so-called EPNJL model, entanglement interaction extended PNJL model. Not to include all these complications in our study and bearing in mind the anticipation that it does not change qualitatively the considered phenomena, we decided to limit ourselves with the NJL model description of the QCD phase diagram.
However, it is in our plans to study the QCD phase diagram with chiral asymmetry in terms of PNJL model but, for now, let us just say that it is possible to show that in the leading order of large-$N_c$ approximation dualities (\ref{16}), (\ref{19}) and (\ref{21}) remain intact even in the PNJL model.

Let us summarize the central results of our paper. (The items (2)-(5) below refer to the case, when one of $\mu_5$ or $\mu_{I5}$ is equal to zero. Whereas in items (6) and (7) the properties of the model are presented under the requirement that both $\mu_5$ and $\mu_{I5}$ are not zero.)

\vspace{0.3cm}

{\bf (1)} It has been demonstrated that there are several duality invariances of the TDP for the massless NJL$_4$ model (1) in the leading order of the large-$N_c$ approximation (see in Sec. \ref{IIIA}). The main duality is a duality correspondence ${\cal D}$ (\ref{16}) between CSB and charged PC phenomena, which was found to take place also in the case of $\nu_{5}\ne 0$ and $\mu_{5}=0$ \cite{ekk,kkz}.
It is a very helpful feature in the context of exploring the phase structure. For example, phase diagrams of Figs 7 and 9 were obtained from other diagrams in different planes using only this duality. Moreover, phase portraits of Figs 10, 11 are self-dual with respect to the transformation $\cal D$ when CSB$\leftrightarrow$charged PC and $\nu\leftrightarrow\nu_5$. The two other  ${\cal D}_{M}$ (\ref{19}) and ${\cal D}_{\Delta}$ (\ref{21}) dualities hold only at the constraints $\Delta=0$ and $M=0$, respectively. As a consequence, if in some cross-section of the general phase diagram the CSB (or the charged PC) phenomenon does not take place at all, then the whole cross-section will be self-dual with respect to the duality transformation ${\cal D}_{\Delta}$ (or ${\cal D}_{M}$), i.e. it is symmetric under interchange $\nu\leftrightarrow\mu_5$ (under interchange $\nu_5\leftrightarrow\mu_5$). For example, the diagram of Fig. 2 is ${\cal D}_{\Delta}$ self-dual, and there the charged PC phase is arranged symmetrically with respect to the line $\nu=\mu_5$. These two dualities ${\cal D}_{M}$ and ${\cal D}_{\Delta}$ in a sense dual to each other with respect to the main duality ${\cal D}$ (see in Fig. 1).

\vspace{0.15cm}

{\bf (2) }It was established in the section \ref{IVB2} that at $\mu_{I5}=0$ and $\mu_{I}=0$ the chiral $\mu_5$ chemical potential is able to catalyze both the CSB and charged PC phenomena. The fact follows from the main duality property ${\cal D}$ of the massless model (1) and means that
at $\nu_5=0$ and $\nu=0$ the TDP (\ref{07}) has two degenerated global minima corresponding to the CSB and charged PC phases, i.e. in the space, filled, e.g., with CSB  phase, a bubble of the charged PC phase (and vice versa) can be created. So, one can observe in  space the mixture (or coexistence) of these two phases, if $\mu_5\ne 0$, $\nu_5=0$, and $\nu=0$.

\vspace{0.15cm}

{\bf  (3)} Applying to the CSB-component of this mixed state the ${\cal D}_{M}$ transformation (\ref{19}), one can obtain the $(\nu_5,\mu)$-phase portrait of the model at $\mu_{5}=0$ and $\mu_{I}=0$, which shows that 
at all values of $\mu_{I5}$  (at $\mu_{5}=0$ and $\mu_{I}=0$) there appears the CSB phase (or symmetric phase at high values of $\mu$). Hence, all the conclusions of the work \cite{braguta2} that there is a catalysis of dynamical CSB by chiral  $\mu_{5}$ chemical potential holds exactly in the same way for chiral isospin $\mu_{I5}$ chemical potential. So there is a catalysis of dynamical CSB by chiral isospin $\mu_{I5}$ chemical potential as well (see in the section \ref{IVB2}). And in order to get all the formulae for $\mu_{I5}$ case one can make the following transformation $\mu_{5}\rightarrow\mu_{I5}$ in the formulae of \cite{braguta2} (firstly, of course, one has to generalize the model to the two flavour case, which is quite trivial).

\vspace{0.15cm}

{\bf  (4)}  According to the constraint duality for PC phenomenon ${\cal D}_{\Delta}$, it was shown that chiral $\mu_5$ chemical potential influences the PC phenomenon in exactly the same way as isospin $\mu_I$ chemical potential. 
It can clearly be demonstrated in the case of  $\mu_{I5}=0$ (see, e.g., in Fig. 2), since the charged PC phase, which can be realized at $\mu_5\ne 0$, $\mu_{I5}=0$ and $\mu_I=0$ is dually-${\cal D}_{\Delta}$ conjugated to the charged PC phase (charged pion condensate, densities etc. are exactly the same) of the model at $\mu_{I}\ne 0$, $\mu_{5}=0$ and $\mu_{I5}=0$  (see at the end of the section \ref{IVB1}).

\vspace{0.15cm}

{\bf  (5)} As it is clear from Fig. 3, chiral $\mu_{5}$ chemical potential alone, in the absence of chiral isospin $\mu_{I5}$ chemical potential, is also able to generate the charged pion condensation in dense quark matter. But this happens for not very extensive regions for $\mu_5$, $\mu_I$ and $\mu$ (see the discussion in Sec. \ref{IVB1}) and for not so large baryon densities.

\vspace{0.15cm}

{\bf  (6)} It was shown that chiral isospin $\mu_{I5}$ chemical potential generates charged pion condensation in dense quark matter (PC$_{d}$ phase) even if isospin $\mu_I$ chemical potential equals to zero (see Figs 10, 11). For this generation to happen one needs to have nonzero chiral chemical potential $\mu_{5}$. In contrast, as it was discussed in \cite{kkz}, this generation requires nonzero values of $\mu_{I}$ in the case of $\mu_{5}=0$, and in the $\mu_{I}=0$ case chiral $\mu_{5}$  chemical potential can take the role of $\mu_{I}$ and allow this generation to happen. This behavior is in accordance with and actually at least can be guessed from 
the constraint duality ${\cal D}_{\Delta}$ (see item (4)).

\vspace{0.15cm}

{\bf  (7)} However, as it is easily seen from the discussion in Sec. \ref{IVC}, in the case, when both types of chiral asymmetry are present in the system (i.e. when both $\mu_5\ne 0$ and $\nu_5\ne 0$), opportunities for the emergence of the charged PC$_d$ phase are greatly extended.
Therefore, for reliable generation of PC$_{d}$ phase it is important to have in the system different chiral imbalances for $u$ and $d$ quarks. \vspace{0.3cm}

As it was discussed in the \cite{kkz} and in Introduction, the dualities akin to ours was obtained in the framework of universality principle (large-$N_{c}$ orbifold equivalence) of phase diagrams in QCD and QCD-like theories in the limit of large $N_{c}$. Are there such dualities in the lattice QCD? We believe that our results can be supported by lattice QCD investigations at least in the case of a zero baryon chemical potential $\mu_B$ (and nonzero isotopic $\mu_I$, or chiral isotopic $\mu_{I5}$, or chiral $\mu_5$ chemical potentials). For example,
some of the phase diagrams at $\mu=0$ can possibly be obtained in lattice simulations and the status of the dualities of the section \ref{IIIA} can be clarified on lattice. Moreover, we hope that our results might shed some new light on phase structure of dense quark matter with isotopic and chiral imbalance and hence could be important for describing physics, for example, in the heavy ion collision experiments, or in an interior of the compact stars.

\appendix{}

\section{Calculation of roots of $P_{\pm}(\eta)$}

In this appendix it will be shown how to get roots of the following quartic equation (general quartic equation could be reduced to the one of this form)
$$
P_{+}(\eta)=\eta^4-2a_{+}\eta^2+b_{+}\eta+c_{+}=0.
$$
First note that we can rewrite it as multiplication of two quadratic equation
$$
(\eta^2+r_{+}\eta+q_{+})(\eta^2-r_{+}\eta+s_{+})=0,
$$
where
$$
-r_+^{2}+q_{+}+s_{+}=-2a_{+},\,\,\,\,\,\,\,\,\,\,\,\,\,q_{+}s_{+}=c_{+},\,\,\,\,\,\,\,\,\,\,\,\,\,\,\,r_{+}s_{+}-r_{+}q_{+}=b_{+}.
$$
From the first and last equations one finds that
$$
q_+=\frac{1}{2} \left(-2 a_{+}+r_+^2-\frac{b_{+}}{r_+}\right),
$$
$$
s_+=\frac{1}{2} \left(-2 a_{+}+r_+^2+\frac{b_{+}}{r_+}\right).
$$
Substituting this into the second equation one gets
that $r_+=\sqrt{R}$, where $R$ is a solution of the following cubic equation
\begin{equation}
X^{3}+A_{+}X=B_{+}X^{2}+C_{+},
\label{cub13}
\end{equation}
where we used notations $A_{+}, B_{+}, C_{+}$ that are given by
$$
A_{+}=
4a_{+}^{2}-c_{+},~~~B_{+}=4a_{+},~~~C_{+}=b_{+}^2.
$$
All three solutions of the cubic equation (\ref{cub13}) are
\begin{equation}
R_{1,2,3}=\frac{1}{3} \left(4 a_{+}+\frac{L_{+}}{\sqrt[3]{J}}+\sqrt[3]{J}\right),\label{A5}
\end{equation}
where
$$
J=\frac{1}{2}(K_{+}+i\sqrt{4 L_{+}^3-K_{+}^2}),~~K_{+}=128 a_{+}^3-36 a_{+} A_{+}+27 b_{+}^2,~~L_{+}=-3A_{+}+16 a_{+}^2,
$$
and $ \sqrt[3]{J}$ in Eq. (\ref{A5}) means each of three possible complex valued roots.
The same can be obtained for $P_{-}(\eta)$ by changing $+\rightarrow-$.


\begin{thebibliography}{999}

\bibitem{Nambu:1961fr}
Y.~Nambu and G.~Jona-Lasinio,  Phys.\ Rev.\  {\bf 122}, 345 (1961);
  Phys.\ Rev.\  {\bf 124}, 246 (1961).

\bibitem{Klevansky:1992qe}
  S.~P.~Klevansky,
  Rev.\ Mod.\ Phys.\  {\bf 64}, 649 (1962).

\bibitem{Hatsuda:1994pi}
  T.~Hatsuda and T.~Kunihiro,
  Phys.\ Rept.\  {\bf 247}, 221  (1994).

\bibitem{Buballa:2003qv}
  M.~Buballa,
  Phys.\ Rept.\  {\bf 407}, 205 (2005).

\bibitem{Fukushima:2003fw}
  K.~Fukushima,
  Phys.\ Lett.\ B {\bf 591}, 277 (2004).

\bibitem{Fukushima:2017csk}
  K.~Fukushima and V.~Skokov,
  Prog.\ Part.\ Nucl.\ Phys.\  {\bf 96}, 154 (2017).

\bibitem{VillavicencioReyes:2004pq}
  C.~L.~Villavicencio Reyes,
  ``Chiral dynamics and pion properties at finite temperature and isospin chemical potential,''
  hep-ph/0510124.

\bibitem{Loewe:2005yn}
  M.~Loewe and C.~Villavicencio,
Phys.\ Rev.\ D {\bf 70}, 074005 (2004);
  Phys.\ Rev.\ D {\bf 71}, 094001 (2005).

\bibitem{Splittorff:2000mm}
  K.~Splittorff, D.~T.~Son and M.~A.~Stephanov,
  Phys.\ Rev.\ D {\bf 64}, 016003 (2001).

\bibitem{Son:2000xc}
  D.~T.~Son and M.~A.~Stephanov,
  Phys.\ Rev.\ Lett.\  {\bf 86}, 592 (2001);
Phys.\ Atom.\ Nucl.\  {\bf 64}, 834 (2001)
   [Yad.\ Fiz.\  {\bf 64}, 899 (2001)].

\bibitem{KogutSinclair}
J. B. Kogut and D. K. Sinclair,
Phys. Rev. D 66, 014508 (2002).

\bibitem{Gupta:2002kp}
  S.~Gupta,
  ``Critical behavior in QCD at finite isovector chemical potential,''  hep-lat/0202005;
O.~Janssen, M.~Kieburg, K.~Splittorff, J.~J.~M.~Verbaarschot and S.~Zafeiropoulos, Phys.\ Rev.\ D {\bf 93}, 094502 (2016).

\bibitem{Brandt:2016zdy}
  B.~B.~Brandt and G.~Endrodi,
  PoS LATTICE {\bf 2016}, 039 (2016);
 B.~B.~Brandt, G.~Endrodi and S.~Schmalzbauer,
  ``The QCD phase diagram for nonzero isospin-asymmetry,''
  arXiv:1712.08190 [hep-lat].

\bibitem{Yakovlev:2004iq}
T. Tatsumi, Progr. Theor. Phys., {\bf 68},  1231 (1982);
T. Takatsuka,  R. Tamagaki,
Progr.Theor. Phys., {\bf 97}, 263 (1997);
  D.~G.~Yakovlev and C.~J.~Pethick,
  Ann.\ Rev.\ Astron.\ Astrophys.\  {\bf 42}, 169 (2004).

\bibitem{son}
J. B. Kogut and D. Toublan, Phys. Rev. D {\bf 64}, 034007
(2001); M. Loewe and C. Villavicencio, Phys. Rev. D {\bf 67}, 074034
(2003); L. He, M. Jin, and P. Zhuang, Phys. Rev. D {\bf 71}, 116001 (2005);
D.~Ebert, K. G.~Klimenko, A. V.~Tyukov and V. C.~Zhukovsky,
  Eur.\ Phys.\ J.\ C {\bf 58}, 57 (2008);
D. C.~Duarte, R. L. S.~Farias and R. O.~Ramos,
  Phys.\ Rev.\  D {\bf 84}, 083525 (2011).

\bibitem{eklim}
D. Ebert and K. G. Klimenko, J.\ Phys.\ G {\bf 32}, 599 (2006);
Eur.\ Phys.\ J.\  C {\bf 46}, 771 (2006).

\bibitem{andersen}
H.~Abuki, R.~Anglani, R.~Gatto, G.~Nardulli and M.~Ruggieri,
  Phys.\ Rev.\  D {\bf 78}, 034034 (2008);
H.~Abuki, R.~Anglani, R.~Gatto, M.~Pellicoro and M.~Ruggieri,
   Phys.\ Rev.\  D {\bf 79}, 034032 (2009);
 R.~Anglani,
  Acta Phys.\ Polon.\ Supp.\  {\bf 3}, 735 (2010).

\bibitem{ak}
J.O.~Andersen and T.~Brauner,
  Phys.\ Rev.\  D {\bf 78}, 014030 (2008);
J.O.~Andersen and L.~Kyllingstad,
 J.\ Phys.\ G {\bf 37}, 015003 (2009);
 Y.~Jiang, K.~Ren, T.~Xia and P.~Zhuang,
Eur.\ Phys.\ J.\ C {\bf 71}, 1822 (2011);
J.~O.~Andersen and P.~Kneschke,
  arXiv:1802.01832 [hep-ph].
%

\bibitem{ekkz}
 D.~Ebert, T. G.~Khunjua, K. G.~Klimenko and V. C.~Zhukovsky,
  Int.\ J.\ Mod.\ Phys.\ A {\bf 27}, 1250162 (2012);
  Phys.\ Atom.\ Nucl.\  {\bf 77}, 795 (2014)
  [Yad.\ Fiz.\  {\bf 77}, 839 (2014)].

\bibitem{gkkz}
  N. V.~Gubina, K. G.~Klimenko, S. G.~Kurbanov and V. C.~Zhukovsky,
  Phys.\ Rev.\ D {\bf 86}, 085011 (2012);
  Phys.\ Atom.\ Nucl.\  {\bf 76}, 1377 (2013)
  [Yad.\ Fiz.\  {\bf 76}, 1443 (2013)].

\bibitem{ekk}
D. Ebert, T. G. Khunjua and K. G. Klimenko, Phys.\ Rev.\ D {\bf 94}, 116016 (2016);
T.~G.~Khunjua, K.~G.~Klimenko, R.~N.~Zhokhov and V.~C.~Zhukovsky,
  Phys.\ Rev.\ D {\bf 95}, no. 10, 105010 (2017).

\bibitem{kkz}
T.~G.~Khunjua, K.~G.~Klimenko and R.~N.~Zhokhov,
Phys. Rev. D 97, 054036  (2018)
  arXiv:1710.09706 [hep-ph] 

\bibitem{liu}
  Y.~Liu and I.~Zahed,
  Phys.\ Rev.\ Lett.\  {\bf 120}, 032001 (2018).

\bibitem{Kharzeev:2007jp}
  D.~E.~Kharzeev, L.~D.~McLerran and H.~J.~Warringa,
  Nucl.\ Phys.\ A {\bf 803}, 227 (2008).

\bibitem{Li:2014bha}
  Q.~Li {\it et al.},
  Nature Phys.\  {\bf 12}, 550 (2016).

\bibitem{Mukherjee:2017bza}
  T.~K.~Mukherjee and S.~Sanyal,
  Mod.\ Phys.\ Lett.\ A {\bf 32}, 1750178 (2017).

\bibitem{Jiang:2014ura}
  Y.~Jiang, X.~G.~Huang and J.~Liao,
  Phys.\ Rev.\ D {\bf 91}, 045001 (2015).

\bibitem{Charbonneau:2009ax}
  J.~Charbonneau and A.~Zhitnitsky,
  JCAP {\bf 1008}, 010 (2010).

\bibitem{CharbonneauHoffman}
J. Charbonneau, K. Hoffman, J. Heyl,
Mon. Not. Roy. Astron. Soc. Lett. {\bf 404},L119 (2010).

\bibitem{andrianov}
 A. A.~Andrianov, D.~Espriu and X.~Planells,
  Eur.\ Phys.\ J.\ C {\bf 73}, 2294 (2013);
 Eur.\ Phys.\ J.\ C {\bf 74}, 2776 (2014);
R.~Gatto and M.~Ruggieri,
  Phys.\ Rev.\ D {\bf 85}, 054013 (2012);
 L.~Yu, H.~Liu and M.~Huang,
 Phys.\ Rev.\ D {\bf 90}, 074009 (2014);
L.~Yu, H.~Liu and M.~Huang,
 Phys.\ Rev.\ D {\bf 94}, 014026 (2016);
G.~Cao and P.~Zhuang,
  Phys.\ Rev.\ D {\bf 92}, 105030 (2015);
 M.~Ruggieri and G.~X.~Peng,
  J.\ Phys.\ G {\bf 43}, no. 12, 125101 (2016).

\bibitem{braguta}
 V.~V.~Braguta, V.~A.~Goy, E.-M.~Ilgenfritz, A.~Y.~Kotov, A.~V.~Molochkov, M.~Muller-Preussker and B.~Petersson,
  JHEP {\bf 1506}, 094 (2015);
V.~V.~Braguta, E.~M.~Ilgenfritz, A.~Y.~Kotov, B.~Petersson and S.~A.~Skinderev,
  Phys.\ Rev.\ D {\bf 93},  034509 (2016).

\bibitem{braguta2}
V. V.~Braguta and A. Y.~Kotov,
 Phys.\ Rev.\ D {\bf 93}, 105025 (2016).

\bibitem{AndrianovEspriu1}
A. Andrianov, V. Andrianov and D. Espriu,
EPJ Web of Conferences {\bf 137}, 01005 (2017).

\bibitem{Farias:2016let}
R.~L.~S.~Farias, D.~C.~Duarte, G.~Krein and R.~O.~Ramos,
  Phys.\ Rev.\ D {\bf 94}, 074011 (2016).

\bibitem{Ruggieri:2011xc}
M.~Ruggieri,
  Phys.\ Rev.\ D {\bf 84}, 014011 (2011).

\bibitem{Ruggieri:2011wd}
  M.~Ruggieri,
  ``Quark Matter with a Chiral Chemical Potential,''
  arXiv:1110.4907 [hep-ph].

\bibitem{Frasca:2016rsi}
  M.~Frasca,
  ``Nonlocal Nambu-Jona-Lasinio model and chiral chemical potential,''
  arXiv:1602.04654 [hep-ph].

\bibitem{Hanada:2011jb}
  M.~Hanada and N.~Yamamoto,
  PoS LATTICE {\bf 2011}, 221 (2011)
  [arXiv:1111.3391 [hep-lat]].

\bibitem{zakharov}
 A.~Avdoshkin, A.~V.~Sadofyev and V.~I.~Zakharov, Phys.\ Rev.\ D {\bf 97}, 085020 (2018); J.~Chao,  arXiv:1808.01928.  

\bibitem{Sannino:2002wp}
  F.~Sannino,
  Phys.\ Rev.\ D {\bf 67}, 054006  (2003).

\bibitem{Lenaghan:2001sd}
  J.~T.~Lenaghan, F.~Sannino and K.~Splittorff,
  Phys.\ Rev.\ D {\bf 65}, 054002 (2002).

\bibitem{Aharony:2007uu}
  O.~Aharony, K.~Peeters, J.~Sonnenschein and M.~Zamaklar,
  JHEP {\bf 0802}, 071 (2008).

\bibitem{Brauner:2016lkh}
  T.~Brauner and X.~G.~Huang,
  Phys.\ Rev.\ D {\bf 94}, 094003 (2016).

\bibitem{Walecka:1974qa} 
  J.~D.~Walecka,
  Annals Phys.\  {\bf 83}, 491 (1974).

\bibitem{Buballa:1996tm} 
  M.~Buballa,
  Nucl.\ Phys.\ A {\bf 611}, 393 (1996);
S.~Carignano, M.~Schramm and M.~Buballa,
  Phys.\ Rev.\ D {\bf 98}, 014033 (2018).

\bibitem{Sakai:2008ga} 
  Y.~Sakai, K.~Kashiwa, H.~Kouno, M.~Matsuzaki and M.~Yahiro,
  Phys.\ Rev.\ D {\bf 78}, 076007 (2008).

\bibitem{Friesen:2014mha} 
  A.~V.~Friesen, Y.~L.~Kalinovsky and V.~D.~Toneev,
  Int.\ J.\ Mod.\ Phys.\ A {\bf 30}, no. 16, 1550089 (2015).

\bibitem{thies}
M.~Thies,
  Phys.\ Rev.\ D {\bf 68}, 047703 (2003).

\bibitem{ekkz2}
D.~Ebert, T. G.~Khunjua, K. G.~Klimenko and V. C.~Zhukovsky,
  Phys.\ Rev.\ D {\bf 90}, 045021 (2014);
  Phys.\ Rev.\ D {\bf 93}, 105022 (2016).

\bibitem{Maldacena:1997re}
  J.~M.~Maldacena,
  Int.\ J.\ Theor.\ Phys.\  {\bf 38}, 1113 (1999)
  [hep-th/9711200].

\bibitem{Hanada:2011ju}
  M.~Hanada and N.~Yamamoto,
  JHEP {\bf 1202}, 138 (2012)
  [arXiv:1103.5480 [hep-ph]].

\bibitem{Kashiwa:2017yvy}
  K.~Kashiwa and A.~Ohnishi,
  Phys.\ Lett.\ B {\bf 772}, 669 (2017)
  [arXiv:1701.04953 [hep-ph]].



\end{thebibliography}
\end{document}